\documentclass[prc,twocolumn,nofootinbib,tightenlines,superscriptaddress,showpacs,preprintnumbers]{revtex4}

\usepackage{amsmath,amssymb,amsfonts,graphicx,subfigure,wrapfig,slashed,color}

\newcommand{\fslash}[1]{\mbox{$\!\not\!#1$}}
\newcommand{\be}{\begin{equation}}
\newcommand{\ee}{\end{equation}}

\begin{document}

\title{
Quark-Jet model for transverse momentum dependent fragmentation functions 
}

\author{W. Bentz}
\email[Corresponding author:~]{bentz@keyaki.cc.u-tokai.ac.jp}
\affiliation{Department of Physics, School of Science, Tokai University,
             4-1-1 Kitakaname, Hiratsuka-shi, Kanagawa 259-1292, Japan}
\affiliation{Radiation Laboratory, Nishina Center, RIKEN, Wako, Saitama 351-0198,
Japan}

\author{A. Kotzinian}
\affiliation{Yerevan Physics Institute, 2 Alikhanyan Brothers Street, 375036 
Yerevan, Armenia}
\affiliation{INFN, Sezione di Torino, 10125 Torino, Italy}

\author{H. H. Matevosyan}
\affiliation{CSSM and ARC Centre of Excellence for Particle Physics at the Tera-Scale,  \\
Department of Physics, University of Adelaide, Adelaide SA 5005, Australia}

\author{Y. Ninomiya}
\affiliation{Department of Physics, School of Science, Tokai University,
             4-1-1 Kitakaname, Hiratsuka-shi, Kanagawa 259-1292, Japan}

\author{A. W. Thomas}
\affiliation{CSSM and ARC Centre of Excellence for Particle Physics at the Tera-Scale,  \\
Department of Physics, University of Adelaide, Adelaide SA 5005, Australia}

\author{K. Yazaki}
\affiliation{Quantum Hadron Physics Laboratory, Nishina Center, RIKEN,
         Wako, Saitama 351-0198, Japan}

\begin{abstract}
In order to describe the hadronization of polarized quarks, we
discuss an extension of the quark-jet model to transverse momentum dependent
fragmentation functions. 
The description is based on a product ansatz, where each factor in the
product represents one of the transverse momentum dependent splitting functions, which 
can be calculated by using effective quark theories.  
The resulting integral equations and sum rules are discussed in detail for the case 
of inclusive pion production. 
In particular, we demonstrate that the 3-dimensional momentum sum rules are satisfied naturally
in this transverse momentum dependent quark-jet model. 
Our results are well suited for numerical
calculations in effective quark theories, and can be implemented 
in Monte-Carlo simulations of polarized quark hadronization processes.

\end{abstract} 


\maketitle

\section{Introduction}
\setcounter{equation}{0}

Quark fragmentation functions (FFs) are key objects for the analysis of inclusive
hadron production in hard scattering processes\cite{FF78}. 
Transverse momentum dependent (TMD) quark FFs, both polarized
and unpolarized, are of particular importance for semi-inclusive
hadron production in $e^+ e^-$ annihilation, semi-inclusive deep inelastic lepton-nucleon
scattering (SIDIS) and proton-proton collisions
\cite{CS81,CS82,CO93,KO95,MT96,BJM9798,BM98,BDR02,BDGMMS07,AR11}.
They are universal, non-perturbative objects, that contain vital information on the
correlation between spin and orbital motion of the fragmenting quark and the
produced hadrons\cite{CM04,JMY05,C13,ESV016}. TMD FFs also are crucial ingredients for accessing the 
TMD parton distribution functions (PDFs) in SIDIS, that encode the 3-dimensional picture
of the nucleon in momentum space\cite{ANS11,SBRS11,ABGMP04,BEMRS15,ABAHMMP15,KPSY16}. 
Particular attention was focused on the
so-called Collins TMD FF \cite{CSS85,CSS88} that allows access to the transversity PDF, the
least well determined of the three leading order PDFs that do not vanish in the collinear
limit. FFs cannot be
calculated in lattice QCD, and therefore effective theories of QCD are very important 
tools to extract information and constraints on TMD FFs.
Important representatives are the quark-jet model\cite{FF78}, the Lund model\cite{AGIS83,AB11}, 
spectator models involving the coupling of quarks to mesons\cite{JMR97,BKMM01,ABM05,BGGM08,MMP10}, and
the Nambu-Jona-Lasinio (NJL) model\cite{NJL} applied in the quark-jet
framework\cite{IBCTY09} using Monte-Carlo techniques\cite{ACY97,B11,MTB11,MTB11a,MTB12,MTB13}.

It is well known\cite{FF78,IBCTY09,MTB11} that a model description of quark FFs
must include the effects of multi-fragmentations in order to reproduce the
main features of the corresponding empirical functions\cite{HKNS07,FSS07,FSEHS15}. 
This is particularly important for the unfavored fragmentation functions, which cannot
be described by assuming one single (elementary) fragmentation step\cite{MTB11,MTB11a}. 
For the 1-dimensional FFs (integrated over
the transverse momentum (TM) of the produced hadron), the quark-jet model of
Field and Feynman\cite{FF78} provides a simple framework to account for multi-fragmentation
processes. It represents a chain of fragmentation processes by
a product of elementary FFs, which can be evaluated in any effective quark theory.
The resulting integral equations of the jet model can be solved directly,
or by using Monte-Carlo methods, which is most convenient if many hadron
channels and resonances are included\cite{MTB11,MTB11a,MTB12,MTB13}. The inclusion of the spin, 
which is directly linked to the transverse momentum dependence, however, remains
a challenging problem for model calculations including multi-fragmentation processes
\cite{AB11,KMT14}. The purpose of this paper
is to provide an analytic framework, based on the assumptions of the successful 
jet model, which can be used for numerical calculations of TMD FFs.
For this, we extend the generalized product ansatz for quark cascades of our
previous work\cite{IBCTY09} to the description of TMD FFs. Limiting ourselves for simplicity
and clarity to the case of inclusive pion production and quark flavor SU(2), we 
derive the explicit forms of the resulting integral equations, and demonstrate 
the validity of the sum rules in the TMD jet model. Our results will allow a
self-consistent formulation of the Monte-Carlo method for polarized quark hadronization,
much needed for the study of various correlations in polarized single - and dihadron
FFs\cite{CCMT12,MKT14,CAD14}.

The outline of the paper is as follows: In Section II we give the operator
definitions of the TMD FFs and discuss their partonic interpretation. 
In Sect. III we derive the integral equations for the TMD FFs from the
basic product ansatz. The explicit forms of the equations will be presented for
the case of inclusive pion production, and the validity of the sum rules
will be confirmed analytically. A summary of our work is given in Section IV.
Further details on the calculations are presented in five Appendices. In
particular, Appendix C presents a list of analytic forms of the elementary
FFs which have been obtained in earlier works
\cite{BKMM01,ABM05,BGGM08,IBCTY09,MMP10} by using effective
quark theories.

The integral equations of the TMD jet model, which we will present in Sect. III.D,
hold in any effective quark theory which does not involve explicit gluon and gauge link 
degrees of freedom, and which satisfies the elementary momentum conservation
and positivity constraints summarized at the end of Sect. III.D.  
The integral equations can then readily be used for numerical calculations. 
It is our hope that our paper
will contribute to a more quantitative understanding of spin dependent fragmentation
processes.

\section{Operator definitions and partonic interpretation}
\setcounter{equation}{0}

The operator definitions of TMD quark FFs follow from the single particle inclusive quark
decay matrix given by\cite{BDR02}
\begin{align}
n_{\beta \alpha} (p_-, {\bold p}_{\perp}; {\bold S}) = \frac{1}{2 z} \int \frac{{\rm d}k_+ {\rm d}k_-}{(2 \pi)^4}
\delta\left(\frac{1}{z} - \frac{k_-}{p_-}\right) N_{\beta \alpha} (p,k; {\bold S})\,,
\label{ndef}
\end{align}
where $z$ is the scaling variable, and the correlator is given by (see Fig.1) 
\footnote{The light-cone components of a 4-vector are defined as
$a^{\mu} = (a^+, a^-, {\bold a}_T)$ with $a^{\pm} = (a^0 \pm a^3)/\sqrt{2}$. Covariant normalization 
is used throughout this paper, and the summation symbol $\sum_n$ in (\ref{Ndef}) includes an
integration over the on-shell momenta $p_n$.} 
\begin{figure}
\begin{center}
\includegraphics[scale=0.55]{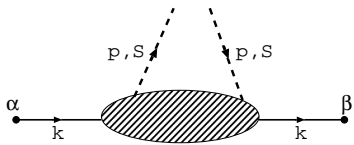}
\caption{Cut diagram representing the correlator of Eq.(\ref{Ndef}). The dots labeled by $\alpha$, $\beta$ indicate
the Dirac indices of the quark field operators, the line labeled by the momentum $k$ represents
the fragmenting quark, and the line labeled by the momentum $p$ and polarization $S$ represents the produced particle.
The shaded oval represents the spectator states $|n\rangle$, and the cut goes through the shaded oval.}
\end{center}
\end{figure}
\begin{align}
N_{\beta \alpha}(p,k; {\bold S}) = {\sum_n} \int {\rm d}^4 \, \omega \, e^{i k \cdot \omega} 
\langle 0| \psi_{\beta}(\omega) | p, n \rangle \langle p, n| \overline{\psi}_{\alpha}(0)|0 \rangle \,.
\label{Ndef}
\end{align}
Here the field operators refer to a given quark flavor ($q=u,d$), which is not indicated
explicitly in this Section, and $k$ and $p$ are the
4-momenta of the fragmenting quark and the produced particle. The state 
\begin{align}
|p, n\rangle = a_h^{\dagger}(p,S) |n \rangle \sqrt{2 p_- \left(2\pi\right)^3}
\label{state}
\end{align}
refers to the produced particle of type $h$ (including isospin) and polarization $S$ 
(which is twice the eigenvalue of the spin operator in 
the direction of ${\bf S}$),  
and a complete set of spectator states
$|n \rangle$. The generic vector ${\bold S}$ specifies the spin 4-vector of the 
produced particle of mass ${\cal M}$ and energy $E_p$ as
\begin{align}
{\cal S}^{\mu} = \left(\frac{{\bold p} \cdot {\bold S}}{\cal M}, \,\,
{\bold S} + \frac{{\bold p} \left({\bold p}\cdot {\bold S}\right)}{{\cal M}\left(E_p+{\cal M}\right)} 
\right) \,.
\label{spin4}
\end{align} 
The operator definitions (\ref{ndef}), (\ref{Ndef}) refer to a frame where
the TM of the produced particle vanishes (${\bold p}_T = 0$) while the fragmenting 
quark has nonzero ${\bold k}_T$.
The vector ${\bold S}$ in (\ref{spin4}) can then be expressed in terms of its transverse components 
${\bold S}_T$ and longitudinal component $S_L$ (helicity) as ${\bold S} = (S_T^1, S_T^2, S_L)$. 
By a transverse Lorentz transformation (see Appendix A for details) one can transform to a frame where the fragmenting 
quark has zero TM (${\bold k}_{\perp} = 0$) and the produced particle has 
${\bold p}_{\perp} = - z {\bold k}_T$, so that we can consider
the decay matrix (\ref{ndef}) as a function of $p_-, {\bold p}_{\perp}$ and ${\bold S}$.

The quark decay matrix (\ref{ndef}) can be expanded in terms of Dirac matrices, with coefficient functions
which are invariant under transverse Lorentz transformations. In leading order, which corresponds to the
limit $p_- \rightarrow \infty$, a set of 4 Dirac matrices ($\Gamma$) contributes to the decay matrix. Their
coefficient functions $\langle \Gamma \rangle \equiv {\rm Tr_D} \left(\Gamma \, n\right)$ can be parametrized 
in terms of 8 FFs in the following way:
\begin{align}
\frac{1}{2 p_-} \langle \gamma^+ \rangle
=  D(z,{\bold p}_{\perp}^2) - \frac{1}{{\cal M}} \epsilon^{ij} k_{T i} S_{Tj}
D_T^{\perp}(z, {\bold p}_{\perp}^2) \,,
\label{vplus} 
\end{align}
\vspace{-0.3cm}
\begin{align}
&\frac{1}{2 p_-} \langle i \sigma^{i +} \gamma_5 \rangle
= S_T^i H_T(z,{\bf p}_{\perp}^2) + \frac{S_L}{{\cal M}} k_T^i 
H_L^{\perp} (z,{\bf p}_{\perp}^2) \nonumber \\
&+ \frac{1}{{\cal M}^2} k_{T}^i  \left({\bf k}_{T}\cdot{\bf S}_T\right) H_T^{\perp}
(z,{\bf p}_{\perp}^2) - \frac{1}{{\cal M}} \epsilon^{ij} k_{T j}  H^{\perp}(z,{\bf p}_{\perp}^2) 
\label{tplusi}\,, 
\end{align}
\vspace{-0.3cm}
\begin{align}
\frac{1}{2 p_-} \langle \gamma^+ \gamma_5 \rangle
= S_L G_L(z,{\bold p}_{\perp}^2) + \frac{1}{{\cal M}} 
\left({\bold k}_{T} \cdot {\bold S}_T \right) G_T (z,{\bold p}_{\perp}^2)\,.
\label{aplus} 
\end{align}
Here $i=1,2$ denote the transverse vector indices, ${\bold k}_T = - {\bold p}_{\perp}/z$, and $\epsilon^{ij} \equiv \epsilon^{-+ij}$ such that $\epsilon^{12}=1$.
The definitions and notations of the 8 leading order FFs in (\ref{vplus}) - (\ref{aplus}) follow
the Trento conventions\cite{BADM04}, except that we assume the large momentum component of the leading produced
particle as $p_- = z k_-$, and we omit the subscript $1$ on all functions because we only consider the leading order here
\footnote{Because the two $T$-odd FFs $D_T^{\perp}$ and $H^{\perp}$ have been introduced first 
in Refs.\cite{MT96} and \cite{CO93}, respectively, they are often called the Mulders-Tangerman
function and the Collins function in the literature. (For the quark distribution functions, their counterparts
are the Sivers function \cite{SI90} and the Boer-Mulders function \cite{BM98}.) The other 6 functions in
(\ref{vplus})-(\ref{aplus}) are $T$-even.}.
 
Next we wish to discuss the partonic interpretation of the various FFs as number densities of the produced
particle ($h$) within a quark, and thereby derive an expression for the ``total fragmentation function'', which will be
used in the next Section to formulate the integral equations of the TMD jet model.
For this purpose we formally define the Dirac matrix valued 4-vector $\Gamma^{\mu}$ as
\begin{align}
\Gamma^{\mu} \equiv \left(\gamma^+, \gamma^+ \gamma^1 \gamma_5,  \gamma^+ \gamma^2 \gamma_5,
\gamma^+ \gamma_5 \right)\,, 
\label{gamma}
\end{align}  
and express the quantities on the left hand sides of Eqs.(\ref{vplus}) - (\ref{aplus}) as
\begin{align}
&\frac{1}{2p_-} \langle \Gamma^{\mu} \rangle \equiv
\frac{1}{2p_-} {\rm Tr}_D \left(\Gamma^{\mu} n(p_-, {\bf p}_{\perp}; {\bf S}) \right)
\nonumber \\
& = \frac{p_-}{2 z}  \int \!{\rm d} \omega^- {\rm d}^2 \omega_T  \, 
 e^{i \left(p_- \omega^- + {\bf p}_{\perp} \cdot {\boldsymbol{\omega}}_T \right)/z} \nonumber \\ 
&\times \langle 0 |  \psi_{\beta}(\omega^-, {\boldsymbol{\omega}}_T) a_h^{\dagger}(p,S) a_h(p,S) 
\overline{\psi}_{\alpha}(0) |0 \rangle \Gamma^{\mu}_{\alpha \beta} 
\label{mat2}  \\
&= \frac{p_-}{z \cdot \sqrt{2}}
\int {\rm d} \omega^- {\rm d}^2 \omega_T \,  
e^{i \left(p_- \omega^- + {\bf p}_{\perp} \cdot {\boldsymbol{\omega}}_T \right)/z}   \nonumber \\
&\times \langle 0 |  \psi_{+ \beta}(\omega^-, {\boldsymbol{\omega}}_T) a_h^{\dagger}(p,S) a_h(p,S) 
{\psi}^{\dagger}_{+ \alpha}(0) |0 \rangle \, \tilde{\Gamma}^{\mu}_{\alpha \beta}\,.
\label{res}
\end{align} 
In the second step we used the relation (\ref{state}) and the completeness of the spectator states $|n \rangle$,
and in the third step we introduced the ``good components'' of the quark field operator by\cite{JJ92,BU96}
\begin{align}
\psi_+ = \frac{1}{\sqrt{2}} \gamma^0 \gamma^+ \psi \equiv \Lambda_{(+)} \psi \,,
\label{plus} 
\end{align}
and defined $\Gamma^{\mu} = \gamma^+ \tilde{\Gamma}^{\mu}$. 
We then introduce the expansion
\begin{align}
\psi_+(\omega^-, {\bf \omega}_T) &= \int \frac{{\rm d}q_-}{\sqrt{2 q_-}}
\frac{{\rm d}^2 q_T}{(2\pi)^{3/2}} \sum_{\lambda} b_{\lambda}(q) 
u_{+ \lambda}(q)  \nonumber \\
&\times e^{-i q_- \omega^-} e^{i {\bf q}_T \cdot {\bf \omega}_T} + \dots \,,
\label{fpsi1} 
\end{align}
where $u_+$ denotes the ``good components'' of the Dirac spinor (see Appendix B for details), and 
the dots ($\dots$) denote the anti-quark terms which do not contribute here. 
Introducing also the quark basis states by
\begin{align}
|{\bf k} \lambda' \rangle = \sqrt{2(2\pi)^3 k_-} \,  b_{\lambda'}^{\dagger}(k) |0 \rangle \,,
\label{quark}
\end{align}
and noting that $\langle {\bf k} \lambda | {\bf k} \lambda \rangle \equiv \langle {\bf k} | {\bf k} \rangle$ is independent of
$\lambda$, we can express (\ref{res}) in a form which is independent of the normalization of states: 
\begin{align}
\frac{1}{2p_-} \langle \Gamma^{\mu} \rangle &= \frac{1}{4} 
\sum_{\lambda' \lambda} \left( \overline{u}_{\lambda'}(k) \Gamma^{\mu} u_{\lambda}(k) \right) \nonumber \\
&\times 
\frac{\langle {\bf k} \lambda | a_h^{\dagger}(p,S)  a_h(p,S) |{\bf k} \lambda' \rangle}
{\langle {\bf k}  | {\bf k}  \rangle} \,.
\label{f2}
\end{align}
In Appendix B we show that the matrix elements in (\ref{f2})
take the form
\begin{align}
\overline{u}_{\lambda'}(k) \Gamma^{\mu} u_{\lambda}(k) = 2 k_- \left( \sigma^{\mu}\right)
_{\lambda' \lambda} \,,
\label{spin}
\end{align}
where we defined $\sigma^{\mu} = \left(1, \boldsymbol{\sigma} \right)$, 
with $\boldsymbol{\sigma} = (\sigma^1, \sigma^2, \sigma^3)$ the usual Pauli matrices.
If we insert (\ref{spin}) into (\ref{f2}) and multiply
both sides by $s_{\mu} \equiv (1, {\bold s})$, where the generic vector ${\bold s}$ 
has Cartesian components $(s_T^1, s_T^2, s_L)$, we obtain\footnote{Like $\Gamma^{\mu}$ and $\sigma^{\mu}$, the quantity
$s^{\mu}$ is {\em not} a Lorentz 4-vector, but Einstein's summation convention still
applies\cite{LP11}.}
\begin{align}
&\frac{1}{2p_-} \langle s_{\mu} \Gamma^{\mu} \rangle \nonumber \\
&= k_- \sum_{\lambda' \lambda} \frac{1}{2} \left({\bf 1} + {\bf s} \cdot \boldsymbol{\sigma}\right) 
_{\lambda' \lambda}  \frac{\langle {\bf k} \lambda | a_h^{\dagger}(p,S)  a_h(p,S) |{\bf k} \lambda' 
\rangle} {\langle {\bf k}  | {\bf k}  \rangle}  \,.  
\label{f3}
\end{align}
Note that in this expression the spin density matrix of the fragmenting quark, 
$\rho({\bf s}) = \frac{1}{2} \left({\bf 1} + {\bold s} \cdot \boldsymbol{\sigma}\right)$, 
appears naturally. Multiplying both sides of (\ref{f3}) by the weight factors
${\rm d} z = {\rm d} p_- / k_-$ and ${\rm d}^2 p_{\perp}$, and expressing the r.h.s. by a
trace operation (Tr), we obtain
\begin{align}
&\frac{1}{2p_-} \langle s_{\mu} \Gamma^{\mu} \rangle \, {\rm d}z \,{\rm d}^2 p_{\perp} \nonumber \\
&= {\rm Tr} \left( \rho({\bf s}) \, \frac{\langle {\bf k} | a_h^{\dagger}(p,S)  a_h(p,S) |{\bf k} 
\rangle} {\langle {\bf k}  | {\bf k}  \rangle} \right) \, {\rm d}p_- \, {\rm d}^2 p_{\perp} \,.
\label{f4}
\end{align}
From this relation it follows that the quantity 
\begin{align}
F(z,{\bf p}_{\perp}; {\bf S}, {\bf s}) \equiv \frac{1}{2p_-} \langle s_{\mu} \Gamma^{\mu} \rangle  
=\frac{1}{2 p_-} s_{\mu} {\rm Tr}_D \left(\Gamma^{\mu} n(p_-, {\bf p}_{\perp}; {\bf S})\right)
\label{f5}
\end{align}
can be interpreted as the number density of the produced particle ($h$) with polarization
${\bf S}$ within the fragmenting quark of polarization ${\bf s}$.    

We can now write down the expression for $F\left(z, {\bf p}_{\perp}; {\bf S}, {\bf s}\right)$,
which follows from the definition (\ref{f5}) and the parametrizations
(\ref{vplus}) - (\ref{aplus}):
\begin{align}
&F\left(z, {\bf p}_{\perp}; {\bf S}, {\bf s}\right)  \nonumber  \\
&= D(z,{\bf p}_{\perp}^2) 
- \frac{1}{\cal M} \left({\bf k}_T \times {\bf S}_T \right)^3 
\, D_T^{\perp}(z,{\bf p}_{\perp}^2) \nonumber \\
&+ \left({\bf s}_T \cdot {\bf S}_T \right) H_T(z,{\bf p}_{\perp}^2)
+ \frac{1}{\cal M} S_L \, \left({\bf k}_T \cdot {\bf s}_T \right) H_L^{\perp}
(z,{\bf p}_{\perp}^2) \nonumber \\
&+ \!\frac{1}{{\cal M}^2} \! \left({\bf S}_T \cdot {\bf k}_T\right) \!
\left({\bf s}_T \cdot {\bf k}_T\right) \! H_T^{\perp} (z,{\bf p}_{\perp}^2) \!  
\nonumber \\
& - \! \frac{1}{\cal M} \! \left( {\bf k}_T \times {\bf s}_T \right)^3 \! 
H^{\perp}(z,{\bf p}_{\perp}^2)  \nonumber \\
&+ \left(S_L \, s_L \right) G_L(z,{\bf p}_{\perp}^2) + 
\frac{1}{\cal M} s_L \left({\bf S}_T \cdot {\bf k}_T \right) G_T(z,{\bf p}_{\perp}^2) \,. 
\label{ftot} 
\end{align}
Here ${\bold k}_T = - {\bold p}_{\perp}/z$, and the superscript $3$ 
denotes the $3$-component of a vector product, i.e., $\left({\bf a}_T \times {\bf b}_T \right)^3
= \epsilon^{ij} a_i b_j$ for any 3-vectors ${\bf a}$ and ${\bf b}$. 
We also remind that the vector ${\bf s}_T$ is transverse to the
momentum of the fragmenting quark, while
${\bf S}_T$ is transverse to the momentum of the produced particle.

In the next Section, we will use the above parametrization for the
``full'' $q \rightarrow {\rm hadron}\, (h)$ FF, which includes effects of 
multi-fragmentation processes,
as well as for the elementary FFs (denoted by small letters $f$, $d$, $d_T^{\perp}$, etc), 
where both $q \rightarrow {\rm hadron} \,(h)$ and $q \rightarrow {\rm quark} \,(Q)$  
processes have to be taken into account
\footnote{Although we used the symbol $h$ (to denote hadron) for the produced particle, the
operator definitions are formally the same for the case where the produced particle is a
quark $(Q)$. For the case of the $q \rightarrow Q$ FFs, the summation 
over $n$ in (\ref{Ndef}) includes the hadronic vacuum state $|0\rangle$.}. (Here $Q=U,D$ denotes
the flavor of a quark in an intermediate state of the fragmentation chain.) 
   
Several sum rules for the full $q \rightarrow h$ function
$F^{(q \rightarrow h)}$ can immediately be derived from the above relations.
Let us for example discuss the
momentum sum rules. Multiplying both sides of (\ref{f4}) by the hadron momentum
${\bf p} \equiv (p_-, {\bf p}_{\perp})$, where $p_- = k_- z$ for fixed $k_-$, and integrating
or summing over all hadronic variables, we obtain
\begin{align}
&\sum_h \, \int_0^1 {\rm d}z \int {\rm d}^2 p_{\perp} \sum_{\pm {\bf S}}
\, \,  {\bf p} \,  F^{(q \rightarrow h)} (z, {\bf p}_{\perp}; {\bf S}, {\bf s}) \nonumber \\
&= {\rm Tr} \, \left(\rho({\bf s}) \, \frac{\langle {\bf k}  | \hat{\bf P} |{\bf k} \rangle} 
{\langle {\bf k} | {\bf k} \rangle} \right) \,\, ,
\label{sumr}
\end{align}
where we defined the momentum operator in terms of hadron variables as
\begin{align}
\hat{\bf P} \equiv \sum_h \, \int_0^{\infty} {\rm d}p_- \, \int {\rm d}^2 p_{\perp} \, 
\sum_{\pm {\bf S}}  \left( {\bf p}\, \, a_h^{\dagger}(p,S) \, a_h(p,S) \right) \,.
\label{momh}
\end{align}
Here and in the following, ${\sum_{\pm {\bf S}}}$ means taking the trace for the spin represented by
${\bf S}$. 
If one allows for an infinite chain of elementary fragmentation processes, the final quark
remainder will have zero longitudinal momentum (LM) fraction, and on average also zero TM: 
$\langle {\bf p}_{\perp} \rangle_{\rm rem} = 0$. 
(We will confirm this point explicitly in the TMD jet model later by using two independent
methods in Sect. III.D. and Appendix E.)  
It then follows that the average value of the hadronic momentum operator $\hat{\bf P}$ 
in the initial quark state is equal to the momentum of the initial quark, which is
${\bf k} = (k_-, {\bf 0}_{\perp})$.  
Eq.(\ref{sumr}) then leads to the LM and TM sum rules\footnote{To derive (\ref{momt}), we use the following identity: 
\begin{align}
\int {\rm d}^2 p_{\perp} \,\,p_{\perp}^i \, p_{\perp}^j\, 
H^{\perp(q \rightarrow h)} (z, {\bf p}_{\perp}^2) 
= \frac{\delta^{ij}}{2} \int {\rm d}^2 p_{\perp} \, {\bf p}_{\perp}^2 \,   
H^{\perp(q \rightarrow h)} (z, {\bf p}_{\perp}^2) \,. 
\nonumber 
\end{align}
Because the TM sum rule (\ref{momt}) has first been introduced in Ref.\cite{ST00}, it is sometimes called the
Sch\"afer - Teryaev sum rule in the literature. 
We note that, although the average TM of the quark remainder after an infinite decay chain
is zero, the magnitude of the fluctuation $\sqrt{\langle {\bf p}^2_{\perp} \rangle_{\rm rem}}$ is nonzero.}
\begin{align}
\sum _h \, \gamma_h \, \int_0^1 {\rm d}z \,z \, \int{\rm d}^2 p_{\perp}  
\, D^{(q \rightarrow h)} (z, {\bf p}^2_{\perp}) = 1 \,,  
\label{moml}
\end{align}
\begin{align}
\sum _h \, \gamma_h \, \int_0^1 \frac{{\rm d}z}{2 z\, M_h} \, \int{\rm d}^2 p_{\perp} \cdot
{\bf p}_{\perp}^2  \, H^{\perp(q \rightarrow h)} (z, {\bf p}_{\perp}^2) = 0 \,,  
\label{momt}
\end{align}
where $\gamma_h$ is the spin degeneracy factor of the hadron and $M_h$ its mass. 
A similar derivation can be given for the $z$ component of the hadronic isospin operator $\hat{T}$, 
which has a form like Eq.(\ref{momh}) with ${\bf p}$ replaced by the $z$ component of the
hadron isospin $t_h$. 
After an infinite decay chain the final quark remainder will have zero average value of isospin 
$z$ component. 
(A simple proof for this is presented in Appendix E.) 
Therefore the average value of $\hat{T}$ in the initial
quark state becomes equal to the isospin $z$ component of the initial quark $\frac{\tau_q}{2}$:
\begin{align}
\sum _h \, \gamma_h \,t_h\, \int_0^1 {\rm d}z  \, \int{\rm d}^2 p_{\perp}  
\, D^{(q \rightarrow h)} (z, {\bf p}^2_{\perp}) = \frac{\tau_q}{2} \,. 
\label{isospin}
\end{align}
The validity of the LM sum rule (\ref{moml}) and the isospin sum rule
(\ref{isospin}) in the quark jet model is well known\cite{FF78,IBCTY09}, and in the
following Section we will also confirm the validity of the TM sum rule
(\ref{momt}).  

\section{Formulation of the TMD jet model}
\setcounter{equation}{0}

In this Section we will formulate the TMD jet model, referring for definiteness and 
simplicity to the case of inclusive pion production. The inclusion of other hadron 
channels is straight forward, in particular if one uses Monte-Carlo methods
\cite{ACY97,B11,MTB11,MTB11a,MTB12,MTB13}.

We first make a few comments on the elementary splitting functions. 
In Appendix C we present model forms of the elementary function $f^{(q \rightarrow Q)}$, 
which is expressed in terms of the 8 
splitting functions $d^{(q \rightarrow Q)}$, $d_T^{\perp (q \rightarrow Q)}$, $\dots$,
$g_T^{ (q \rightarrow Q)}$ similar to Eq. (\ref{ftot}), and the elementary 
function $f^{(q \rightarrow \pi)}$, for which only the
spin independent term $d^{(q \rightarrow \pi)}$ and the quark - spin dependent 
term $\propto h^{\perp (q \rightarrow \pi)}$ contribute. These forms, which are obtained in 
any effective theory which
involves the coupling of constituent quarks to pions, are given in lowest order of the pion-quark
coupling constant, i.e., the tree diagrams for the $T$-even functions (see Fig. 2 of Appendix C) 
and the one-loop graphs for the $T$-odd functions (see Figs.3 and 4 of Appendix C).  
One peculiar feature of those functions is that
the virtual quark can fragment into an on-shell quark and a real pion only with a 
certain probability $1-Z_Q$, which is actually equal to the probability to find a 
constituent quark with its virtual pion cloud\cite{IBCTY09,MMP10}. 
(Typical values are $Z_Q \simeq 0.8$.) 
More precisely, the elementary $q \rightarrow Q$ FF 
can be expressed in the form
\begin{align}
&f^{(q \rightarrow Q)} \left(z, {\bf p}_{\perp}; {\bf S}, {\bf s}\right) 
\equiv  Z_Q \delta(1-z) \delta^{(2)}({\bf p}_{\perp}) \, \delta(\tau_{Q}, \tau_{q})  \nonumber \\  
&\times \frac{1}{2}\left(1 + {\bf S} \cdot {\bf s} \right) + (1-Z_Q) \hat{f}^{(q \rightarrow Q)} 
\left(z, {\bf p}_{\perp}; {\bf S}, {\bf s} \right) \,,
\label{fg}
\end{align}
where the first term involves the probability $Z_Q$ that the quark does not fragment at all
\footnote{The spin structure of the non-fragmentation term is explained in Appendix C.
In practice, this term only serves to renormalize the elementary fragmentation
functions, as explained in Appendix D.}
, and accordingly the new function $\hat{f}^{(q\rightarrow Q)}$ is normalized to $1$:
\begin{align}
&\int_0^1 {\rm d} z \, \int {\rm d}^2 p_{\perp} \sum_{\pm {\bf S}} \sum_{\tau_Q}
\hat{f}^{(q \rightarrow Q)}(z, {\bf p}_{\perp}; {\bf S}, {\bf s}) = 1 \,.
\label{fnorm}  
\end{align}
This renormalized elementary function $\hat{f}^{(q\rightarrow Q)}$ is again parametrized
as in Eq. (\ref{ftot}) in terms of the 8 
splitting functions $\hat{d}^{(q \rightarrow Q)}$, $\hat{d}_T^{\perp (q \rightarrow Q)}$, $\dots$,
$\hat{g}_T^{ (q \rightarrow Q)}$. (Explicit model forms obtained in lowest order perturbation theory
are given in Appendix C.) 

For the formulation of the product ansatz, it will be convenient to define the elementary 
$q \rightarrow Q$ FF for the case where the incoming quark ($q$) has polarization ${\bf s}$ and the outgoing 
quark ($Q$) is unpolarized:
\begin{align}
&\hat{f}^{(q \rightarrow Q)}(z, {\bf p}_{\perp}; {\bf s}) \equiv
\sum_{\pm {\bf S}} \, \hat{f}^{(q \rightarrow Q)}(z, {\bf p}_{\perp}; {\bf S}, {\bf s}) 
\nonumber \\
&= 2 \left[\hat{d}^{(q\rightarrow Q)} (z, {\bf p}_{\perp}^2) + \frac{1}{M z} \left({\bf p}_{\perp} \times {\bf s}_T\right)^3
\, \hat{h}^{\perp (q \rightarrow Q)} (z, {\bf p}_{\perp}^2) \right]\,, 
\label{qqunpol}
\end{align}
where $M$ is the constituent quark mass. 
The renormalized elementary $q \rightarrow \pi$ FF is related to the above function by (see Refs.\cite{MMP10,IBCTY09})
\footnote{We denote
$\tau_q = (1, -1)$ for $(u,d)$ and $\tau_{\pi} = (1,0,-1)$ for $(\pi^+, \pi^0, \pi^-)$.}
\begin{align}
\hat{f}^{(q \rightarrow \pi)}(z, {\bf p}_{\perp}; {\bf s}) = \hat{f}^{(q \rightarrow Q)}(1-z, - {\bf p}_{\perp}; {\bf s})
|_{\tau_Q = \tau_q - 2 \tau_{\pi}} \, ,
\label{relation} 
\end{align}
and is normalized to $1$ according to (\ref{fnorm}).
For later reference, we finally note that from (\ref{fnorm}) the quark renormalization factor 
is expressed in terms
of the unrenormalized integrated $q \rightarrow Q$ FF $d^{(q \rightarrow Q)}(z)$ as follows:
\begin{align}
1-Z_Q = 2 \sum_{\tau_Q} \int_0^1 {\rm d} z \, d^{(q \rightarrow Q)}(z) \,.
\label{zform}
\end{align}

\subsection{Product ansatz}

In order to describe multistep fragmentation (quark cascade) processes, in our previous work\cite{IBCTY09} 
we expressed the integrated $q \rightarrow \pi$ FF by a sum of products of elementary
$q \rightarrow Q$ FFs, introducing the maximum number of pions $(N)$ which can be produced by the
fragmenting quark. It was shown that the momentum and isospin sum rules
are satisfied only in the limit of $N\rightarrow \infty$
\footnote{Although this indicates a conceptual limitation of the jet model, which arises from several assumptions 
like scaling, leading twist and factorization, we take the limit $N\rightarrow \infty$ here, because
one of the purposes of this paper is just to demonstrate the validity of the sum rules in this limit
for the TMD case.}. In this limit one recovers
the original jet model of Field and Feynman\cite{FF78}, where the FF is expressed from the
start by an infinite product of renormalized $q \rightarrow Q$ FFs, corresponding to our
quantity $\hat{f}^{(q \rightarrow Q)}$ of Eq.(\ref{fg}). In Appendix D we show that the same
line of argument can be used also for the TMD case, i.e., the first 
(non-fragmentation) term of (\ref{fg}) can be processed so as to express the full $q \rightarrow \pi$ 
FF in terms of 
products of the renormalized elementary $q \rightarrow Q$ FFs of (\ref{fg}). In order to keep the
formulas of the main part as simple as possible, we use the limit $N \rightarrow \infty$ from the start here.
We will use the 
following notations for multi-dimensional momentum integrations:
\begin{align}
\int {\cal D}^N \eta &\equiv \int_0^1 {\rm d}\eta_1 \,\int_0^1 {\rm d}\eta_2 \cdots \int_0^1 {\rm d}\eta_N 
\,\,,\nonumber \\
\int {\cal D}^{2N} p_{\perp} &\equiv \int {\rm d}^2 p_{1 \perp} \,  \int {\rm d}^2 p_{2 \perp} \cdots \int {\rm d}^2 p_{N \perp}
\,\,. 
\label{multi} 
\end{align} 
The product ansatz is then as follows:
\begin{align}
&F^{(q \rightarrow \pi)}(z, {\bf p}_{\perp}; {\bf s}) 
= {\rm lim}_{N\rightarrow \infty} \, \sum_{m=1}^{N} \, \int {\cal D}^N \eta \, \int {\cal D}^{2N} p_{\perp} \, 
\sum_{\tau_{Q_N}}    \nonumber \\
&\times \hat{f}^{(q \rightarrow Q_1)}(\eta_1, {\bf p}_{1 \perp}; {\bf s}) \,  
\hat{f}^{(Q_1 \rightarrow Q_2)}(\eta_2, {\bf p}_{2 \perp} - \eta_2 {\bf p}_{1 \perp}; \langle {\bf S}_1 \rangle)
\nonumber \\
&\times \cdots \times 
\hat{f}^{(Q_{N-1} \rightarrow Q_N)}(\eta_N, {\bf p}_{N \perp} - \eta_N {\bf p}_{N-1 \perp}; \langle {\bf S}_{N-1} \rangle)
\nonumber \\ 
&\times \delta(z - z_m) \, \delta^{(2)}\left({\bf p}_{\perp} - ({\bf p}_{m-1 \perp} -{\bf p}_{m \perp})\right)
\nonumber \\
&\times \delta\left(\tau_{\pi}, (\tau_{Q_{m-1}} - \tau_{Q_m}) /2\right) 
\equiv {\rm lim}_{N \rightarrow \infty} \sum_{m=1}^N F_m^{(q \rightarrow \pi)}(z, {\bf p}_{\perp}; {\bf s}) \,.   
\label{ansatz}
\end{align}
Here the function $\hat{f}^{(q \rightarrow Q_1)}(\eta_1, {\bf p}_{1 \perp}; {\bf s})$ is the elementary
FF for the first step, which refers to the case where the incoming quark $(q)$ has polarization ${\bf s}$ and no 
TM (${\bf k}_{\perp} = 0$), and the outgoing quark $(Q_1)$ is unpolarized and has momentum variables
$\left(\eta_1, {\bf p}_{1 \perp}\right)$. 
The function 
$\hat{f}^{(Q_i \rightarrow Q_j)}(\eta_j, {\bf p}_{j \perp} - \eta_j {\bf p}_{i \perp}; \langle {\bf S}_i \rangle)$
for the $j$th step refers to the case where the incoming quark $(Q_i)$ has momentum variables
$\left(\eta_i, {\bf p}_{i \perp}\right)$ and a polarization $\langle {\bf S}_i \rangle$,
which is defined as the mean polarization density of the outgoing quark of the $i$th step (which depends implicitly
on the momentum variables of the steps $1,2, \dots i$), while the outgoing quark $(Q_j)$ has momentum
variables $\left(\eta_j, {\bf p}_{j \perp}\right)$ and its spin is not observed.
In (\ref{ansatz}) we applied the rule
(\ref{general}) for making a transverse Lorentz transformation in each step of the fragmentation chain.
The delta functions in (\ref{ansatz}) select a meson which is produced in the
m-th step with LM fraction $z_m$ of the initial quark, where
\begin{eqnarray}
z_m = \eta_1 \eta_2 \cdots \eta_{m-1} \cdot (1-\eta_m)  \label{zm}
\end{eqnarray}
for $m>1$, and $z_1 = 1 - \eta_1$ for $m=1$.
In (\ref{ansatz}) a sum over repeated quark flavor indices is implied, 
and for $m=1$ we define 
${\bf p}_{0 \perp} \equiv {\bf k}_{\perp} = 0$ and ${\bf S}_0 = {\bf s}$. 

The main difference to the case of the integrated FFs\cite{IBCTY09} is the 
spin structure of the product ansatz (\ref{ansatz}), which will be explained
in the following Subsection.  

\subsection{Spin structure of the product ansatz}

Here we wish to explain the spin structure of the product ansatz
(\ref{ansatz}). For this purpose, we keep 
only the spin variables in most parts of this Subsection, suppressing momentum and isospin labels for simplicity. 

Because the $q \rightarrow \pi$ FF is obtained from  
a chain of elementary fragmentation processes, averaging over the spin of the final quark remainder, 
we express it formally as
\begin{align}
F({\bf s}) = \lim_{N \rightarrow \infty} \, {\rm Tr}
\left[ \left(a^* + {\bf b}^* \cdot {\boldsymbol{\sigma}} \right)^N \, \rho({\bf s}) \,
\left(a + {\bf b} \cdot {\boldsymbol{\sigma}} \right)^N \right] \,.   
\label{form}
\end{align}
Here ${\rm Tr}$ denotes the trace of a spin $2 \times 2$ matrix, 
$\rho({\bf s})$ is the spin density matrix of the initial quark as before, 
and in order to avoid long expressions for products 
we use the symbolic notations
\begin{align}
&\left(a + {\bf b}\cdot \boldsymbol{\sigma}\right)^n \nonumber \\
&\equiv
\left(a_1 + {\bf b}_1\cdot \boldsymbol{\sigma}\right) \cdot 
\left(a_2 + {\bf b}_2\cdot \boldsymbol{\sigma}\right) \cdot \dots \cdot
\left(a_n + {\bf b}_n\cdot \boldsymbol{\sigma}\right) \,,
\label{def1}  \\
&\left(a^* + {\bf b}^*\cdot \boldsymbol{\sigma}\right)^n \nonumber \\
&\equiv
\left(a^*_n + {\bf b}^*_n\cdot \boldsymbol{\sigma}\right) \cdot \dots
\cdot  
\left(a^*_2 + {\bf b}^*_2\cdot \boldsymbol{\sigma}\right) \cdot
\left(a^*_1 + {\bf b}^*_1\cdot \boldsymbol{\sigma}\right) \,,
\label{def2}
\end{align}
where $a_n$ and ${\bf b}_n$ depend on the momentum variables of the $n$th fragmentation step. 

Our aim is to express (\ref{form}) as a product of $N$ factors. 
For this, we first note that the matrix corresponding to the first fragmentation step $(q \rightarrow Q_1)$ 
can be expressed as
\begin{align}
\tilde{f}_1({\bf s}) &= \left(a_1^* + {\bf b}_1^* \cdot {\boldsymbol{\sigma}} \right) \, \rho({\bf s}) \,
\left(a_1 + {\bf b}_1 \cdot {\boldsymbol{\sigma}} \right) \nonumber \\
& \equiv \frac{1}{2}
\left(f_1({\bf s}) + {\boldsymbol{\sigma}} \cdot {\bf f}_1({\bf s}) \right)
\label{first1}  \\ 
&= f_1({\bf s}) \, \rho\left(\langle {\bf S}_1 \rangle \right) \,, 
\label{first}
\end{align}
where in (\ref{first1}) we defined the functions
\begin{align}
&f_1({\bf s}) = {\rm Tr} \left[ \left(a_1^* + {\bf b}_1^* \cdot \boldsymbol{\sigma} 
\right) \, \rho({\bf s}) \,  
\left(a_1 + {\bf b}_1 \cdot \boldsymbol{\sigma} \right) \right]
\label{f1s}\,, \\
&{\bf f}_1({\bf s}) =  {\rm Tr} \left[ \left(a_1^* + {\bf b}_1^* \cdot \boldsymbol{\sigma} 
\right) \, \rho({\bf s}) \,  
\left(a_1 + {\bf b}_1 \cdot \boldsymbol{\sigma} \right) {\boldsymbol{\sigma}} \right] \,,
\label{f1v} 
\end{align}   
while in (\ref{first}) we used the spin density matrix  
${\displaystyle \rho(\langle {\bf S}_1 \rangle) = \frac{1}{2} \left(1 + \langle {\bf S}_1 \rangle \cdot 
\boldsymbol{\sigma} \right)}$, where 
\begin{align}
\langle {\bf S}_1 \rangle = \frac{{\bf f}_1({\bf s})}{f_1({\bf s})}
\label{s1av}
\end{align}
is the average polarization density of $Q_1$ (after the first step). Because of
$|\langle {\bf S}_1 \rangle|\leq 1$, the quark $Q_1$ is in a partially polarized state.

The matrix corresponding to the first and second fragmentation steps
($q \rightarrow Q_1 \rightarrow Q_2$) can then be expressed as 
\begin{align}
\tilde{f}_2({\bf s}) &=  \left(a_2^* + {\bf b}_2^* \cdot {\boldsymbol{\sigma}} \right)
\, f_1({\bf s}) \, \rho\left(\langle {\bf S}_1 \rangle \right)  
\left(a_2 + {\bf b}_2 \cdot {\boldsymbol{\sigma}} \right)   \nonumber \\
&\equiv f_1({\bf s}) \, \frac{1}{2} \left(f_2(\langle {\bf S}_1 \rangle) + {\boldsymbol{\sigma}} \cdot 
{\bf f}_2(\langle {\bf S}_1 \rangle) \right)
\label{second1} \\
&= f_1({\bf s}) \, f_2(\langle {\bf S}_1 \rangle) \, \rho\left(\langle {\bf S}_2 \rangle \right)  \,, 
\label{second}
\end{align}
where in (\ref{second1}) we defined the functions
\begin{align}
&f_2(\langle {\bf S}_1\rangle) = {\rm Tr} \left[ \left(a_2^* + {\bf b}_2^* \cdot \boldsymbol{\sigma} 
\right) \, \rho(\langle {\bf S}_1 \rangle) \,  
\left(a_2 + {\bf b}_2 \cdot \boldsymbol{\sigma} \right) \right]
\label{f2s}\,, \\
&{\bf f}_2(\langle {\bf S}_1 \rangle)  = {\rm Tr} \left[ \left(a_2^* + {\bf b}_2^* \cdot \boldsymbol{\sigma} 
\right) \, \rho(\langle {\bf S}_1 \rangle) \,  
\left(a_2 + {\bf b}_2 \cdot \boldsymbol{\sigma} \right) {\boldsymbol{\sigma}} \right] \,,
\label{f2v} 
\end{align}  
while in (\ref{second}) we used the spin density matrix  
${\displaystyle \rho(\langle {\bf S}_2 \rangle) = \frac{1}{2} \left(1 + \langle {\bf S}_2 \rangle \cdot 
\boldsymbol{\sigma} \right)}$, where 
\begin{align}
\langle {\bf S}_2 \rangle = \frac{{\bf f}_2(\langle {\bf S}_1 \rangle)}{f_2(\langle {\bf S}_1 \rangle)}
\label{s2av}
\end{align}
is the average polarization density of $Q_2$ (after the second step). 

We can continue in this way, and after $N$ steps we obtain for the FF (\ref{form}) 
\begin{align}
F({\bf s}) &= \lim_{N \rightarrow \infty} \, f_1({\bf s}) \, f_2(\langle {\bf S}_1 \rangle) \dots 
f_N(\langle {\bf S}_{N-1} \rangle) \, 
{\rm Tr} \, \rho \left(\langle {\bf S}_N \rangle \right)   \nonumber \\    
&= \lim_{N \rightarrow \infty} \, f_1({\bf s}) \, f_2(\langle {\bf S}_1 \rangle) \dots 
f_N(\langle {\bf S}_{N-1} \rangle) \,.
\label{desir}
\end{align}
Eq.(\ref{desir}) is the desired result, because it expresses the quantity (\ref{form}) 
by a product of $N$ factors, where each factor is given in terms of the elementary
FF. This concludes the derivation of the spin structure of the product ansatz (\ref{ansatz}).

We finally comment on the relation between the matrix representation of the elementary FFs
used in this Subsection, and the form (\ref{ftot}).
For definiteness we consider the FF for the first step, which in Eq.(\ref{first1}) 
was expressed in spin matrix form
as ${\displaystyle \tilde{f}_1({\bf s}) = \frac{1}{2} \left(f_1({\bf s}) + {\boldsymbol{\sigma}} 
\cdot {\bf f}_1({\bf s}) \right)}$. The connection to the form (\ref{ftot}) for the elementary 
$q \rightarrow Q_1$ case is given by 
\begin{align}
f_1({\bf S}_1, {\bf s}) = {\rm Tr} \left( \tilde{f}_1({\bf s}) \rho({\bf S}_1) \right)
= \frac{1}{2} \left(f_1({\bf s}) + {\bf S}_1 \cdot {\bf f}_1({\bf s}) \right) \,,
\label{eend}
\end{align}
where again the subscript $1$ on the functions $f$ and ${\bf f}$ is used to denote the dependence 
on the momentum variables for the
first step. In (\ref{eend}), ${\bf S}_1$ is considered 
simply as an auxiliary variable, i.e., if one knows $f_1({\bf S}_1, {\bf s})$ as a function
of ${\bf S}_1$, one also knows the
matrix valued function $\tilde{f}_1({\bf s})$. (We note that an analogous trace operation was performed in
(\ref{f4}) for the initial quark.) 
Eq.(\ref{eend}) also provides a natural extension of the formalism in Sect. II, where
the polarization $S$ in (\ref{state}) implicitly referred to a fully polarized state,
to the case of partial polarization.  

Returning to the full notations including the momentum
and isospin variables, comparison of (\ref{ftot}) with (\ref{eend}) gives 
\begin{align}
&\hat{f}^{(q \rightarrow Q_1)}(\eta_1,{\bf p}_{1 \perp}; {\bf s})  
= 2 \left[\hat{d}^{(q\rightarrow Q_1)} (\eta_1, {\bf p}_{1 \perp}^2) \right. \nonumber \\
&\left. + \frac{1}{M \eta_1} \left({\bf p}_{1 \perp} 
\times {\bf s}_T\right)^3 \, \hat{h}^{\perp (q\rightarrow Q_1)} (\eta_1, {\bf p}_{1 \perp}^2) \right]  
\label{qqunpol1}
\end{align} 
in agreement with (\ref{qqunpol}), and
\footnote{Eq.(\ref{sav1}) shows only the transverse part of $\hat{{\bf f}}^{(q \rightarrow Q_1)}$
without the contribution from the last term $\propto s_L$ in the elementary version of Eq.(\ref{ftot}).
It will become clear in Subsection III.D. that this term does not contribute to inclusive pion
production. Also, there is a longitudinal part of 
$\hat{{\bf f}}^{(q \rightarrow Q_1)}$ which arises from the terms $\propto S_L$ in the elementary
version of (\ref{ftot}). Because the total FF for $q \rightarrow \pi$ consists only of the 
unpolarized $(D)$ and the Collins $(H^{\perp})$ terms of (\ref{ftot}), this part
does not contribute either.}  
\begin{align}
&\hat{{\bf f}}^{(q \rightarrow Q_1)}(\eta_1, {\bf p}_{1 \perp}; {\bf s})  \nonumber \\ 
&= 2 \left[  \frac{1}{M \eta_1} {\bf p}'_{1 \perp} \, \, \hat{d}_T^{\perp (q\rightarrow Q_1)}(\eta_1, {\bf p}_{1 \perp}^2) 
+ {\bf s}_T \hat{h}_T^{(q \rightarrow Q_1)}(\eta_1, {\bf p}_{1 \perp}^2) \right. \nonumber \\
&\left. + \frac{1}{M^2 \eta_1^2} \, {\bf p}_{1 \perp} \left({\bf s}_T \cdot {\bf p}_{1 \perp} \right) 
\hat{h}_T^{\perp (q \rightarrow Q_1)}(\eta_1, {\bf p}_{1 \perp}^2) \right] \,.   
\label{sav1}
\end{align}
If ${\bf p}_{1 \perp} = (p_{1 \perp}^1, p_{1 \perp}^2)$, the vector ${\bf p}_{1 \perp}'$ is defined by
${\bf p}_{1 \perp}' = (- p_{1 \perp}^2, p_{1 \perp}^1)$. To get the corresponding functions for the
second step, one has to replace the momentum variables $(\eta_1, {\bf p}_{1 \perp})$ by
$(\eta_2, {\bf p}_{2 \perp} - \eta_2 {\bf p}_{1 \perp})$, while according to (\ref{second}) the spin variable 
${\bf s}$ should be replaced by $\langle {\bf S}_1 \rangle$, which is the ratio of the 2 functions 
given above for the first step.

\subsection{Integral equations}

Let us now proceed with the product ansatz (\ref{ansatz}) to derive the integral
equation for the FF in the TMD jet model. For a fixed $m$ in (\ref{ansatz}), we can integrate over the
variables $\eta_k$, ${\bf p}_{k \perp}$ for $k>m$ using the normalization (\ref{fnorm}). The integrations
over $\eta_m$, ${\bf p}_{m \perp}$ are then performed by using the delta functions. Making a shift
$\eta_m \rightarrow 1 - \eta_m$ and using (\ref{relation}), the result of
these integrations is
\begin{align}
&\sum_{\tau_{Q_m}}\, \int_0^1 {\rm d} \eta_m \, \int {\rm d}^2 p_{m \perp} \,  
\,\delta(z - z_m) \nonumber \\ 
&\times \hat{f}^{(Q_{m-1} \rightarrow Q_m)} (\eta_m, {\bf p}_{m \perp} - \eta_m {\bf p}_{m-1 \perp} ; 
\langle {\bf S}_{m-1} \rangle) \nonumber \\
&\times \delta({\bf p}_{\perp} - ({\bf p}_{m-1 \perp} - {\bf p}_{m \perp}))
\, \delta(\tau_{\pi}, (\tau_{Q_{m-1}} - \tau_{Q_m})/2) 
\nonumber \\
&= \int_0^1 {\rm d} \eta_m \,\delta(z - \eta_1 \eta_2 \dots \eta_m)   \nonumber \\  
&\times \hat{f}^{(Q_{m-1} \rightarrow \pi)} (\eta_m, {\bf p}_{\perp} - \eta_m {\bf p}_{m-1 \perp}; 
\langle {\bf S}_{m-1} \rangle ) \,.
\label{intm}
\end{align}
In this way, the function $F_m^{(q \rightarrow \pi)}$ of Eq.(\ref{ansatz}) becomes
\begin{align}
&F_m^{(q \rightarrow \pi)}(z, {\bf p}_{\perp}; {\bf s}) 
=  \int {\cal D}^m \eta \, \int {\cal D}^{2(m-1)} p_{\perp}  \nonumber \\ 
&\times \hat{f}^{(q \rightarrow Q_1)}(\eta_1, {\bf p}_{1 \perp}; {\bf s}) 
\, \hat{f}^{(Q_1 \rightarrow Q_2)}(\eta_2, {\bf p}_{2 \perp} - \eta_2 {\bf p}_{1 \perp}; \langle {\bf S}_1 \rangle )
\cdots  \nonumber \\
&\times \hat{f}^{(Q_{m-2} \rightarrow Q_{m-1})}(\eta_{m-1}, {\bf p}_{m-1 \perp}\! -\!\eta_{m-1} {\bf p}_{m-2 \perp}; 
\langle {\bf S}_{m-2} \rangle )  \nonumber \\ 
&\times \hat{f}^{(Q_{m-1} \rightarrow \pi)} (\eta_m, {\bf p}_{\perp}\! - \! \eta_m {\bf p}_{m-1 \perp}; 
\langle {\bf S}_{m-1} \rangle )
\delta(z - \eta_1 \eta_2 \cdots \eta_m) \,. 
\label{ansatz2}
\end{align}
In order to obtain a recursion relation for the functions $F_m^{(q\rightarrow \pi)}$, we carry out the steps
explained in Appendix D (see Eqs.(\ref{trans}) - (\ref{neq01})), and obtain for $m>1$ 
\begin{align}
&F_m^{(q \rightarrow \pi)} (z, {\bf p}_{\perp}; {\bf s}) 
= \int {\cal D}^2 \eta \, \int {\cal D}^{4} p_{\perp} \,  \nonumber \\
&\! \times\! \delta(z - \eta_1 \eta_2) \delta^{(2)}({\bf p}_{\perp} - {\bf p}_{2 \perp} - \eta_2 {\bf p}_{1 \perp})
\hat{f}^{(q \rightarrow Q)}(\eta_1, {\bf p}_{1 \perp} ; {\bf s}) \nonumber \\
&\times F_{m-1}^{(Q \rightarrow \pi)} (\eta_2, {\bf p}_{2 \perp}; \langle {\bf S}_1 \rangle ) \,,   
\label{recurs1}
\end{align}
where $\langle {\bf S}_1 \rangle$ is the mean polarization density of the quark produced in the first step
and depends on the momentum variables $\left(\eta_1, {\bf p}_{1 \perp} \right)$ (for the explicit form, see
Eq.(\ref{sav}) of the following Subsection), while for $m=1$ we have
\begin{align}
F_1^{(q \rightarrow \pi)} (z, {\bf p}_{\perp}; {\bf s})
= \hat{f}^{(q \rightarrow \pi)} (z, {\bf p}_{\perp}; {\bf s}) \,.
\label{neq1}
\end{align}
Because the total FF is obtained by performing the sum over $m$ and taking the limit $N \rightarrow \infty$ 
(see (\ref{ansatz})), 
it satisfies the following integral equation:
\begin{align}
&F^{(q\rightarrow \pi)}(z,{\bf p}_{\perp}; {\bf s}) = \hat{f}^{(q\rightarrow \pi)}(z,{\bf p}_{\perp}; {\bf s})
\nonumber \\
&+ \int {\cal D}^2 \eta \, \int {\cal D}^{4} p_{\perp} \, \delta(z - \eta_1 \eta_2) \nonumber \\
&\times \delta^{(2)}({\bf p}_{\perp} - {\bf p}_{2 \perp} - \eta_2 {\bf p}_{1 \perp})
\hat{f}^{(q \rightarrow Q)}(\eta_1, {\bf p}_{1 \perp} ; {\bf s} ) \nonumber \\
&\times F^{(Q \rightarrow \pi)} (\eta_2, {\bf p}_{2 \perp}; \langle {\bf S}_1 \rangle) \,.   
\label{recurs}
\end{align}
More explicit forms of this integral equation will be derived in the following Subsection.
Here we add remarks on the following two points: 
First, the SU(2) flavor dependence of all $q \rightarrow \pi$ and 
$q \rightarrow Q$ FFs in this paper (elementary or full) can be expressed by
\begin{align}
Z^{(q \rightarrow \pi)} &= 
\frac{1}{3} Z_{(0)}^{(q \rightarrow \pi)}
+ \frac{1}{2} \tau_q \tau_{\pi} Z_{(1)}^{(q \rightarrow \pi)} \,,
\label{isivqp} \\
Z^{(q \rightarrow Q)} &= 
\frac{1}{2} Z_{(0)}^{(q \rightarrow Q)}
+ \frac{1}{2} \tau_q \tau_{Q} Z_{(1)}^{(q \rightarrow Q)} \,.
\label{isivqq}
\end{align}
Here $Z = \hat{f}$ for the elementary functions, and $Z = F$ for the full functions, 
and the subscripts $(0)$ and $(1)$
denote the isoscalar and isovector parts\footnote{For the isoscalar and isovector functions
$Z_{(\alpha)}$, the
distinction between the quark labels $q$ and $Q$ is irrelevant.}.
These definitions are convenient for the discussion of sum rules because of the following relations:
\begin{align}
&\sum_{\tau_{\pi}} Z^{(q \rightarrow \pi)} = Z_{(0)}^{(q \rightarrow \pi)}\,,\,\,\,\,\,\,\,\,
\sum_{\tau_{\pi}} \tau_{\pi} Z^{(q \rightarrow \pi)} = \tau_q Z_{(1)}^{(q \rightarrow \pi)}  \,.
\label{isiv}
\end{align}
By using the forms (\ref{isivqp}) and (\ref{isivqq}) in the integral equation (\ref{recurs}),
the sum over the intermediate quark flavors can be easily carried out, and one obtains
two separate integral equations, of the same form as the original equation (\ref{recurs}),
for the isoscalar ($\alpha=0$) and isovector ($\alpha=1$) parts:
\begin{align}
&F_{(\alpha)}^{(q\rightarrow \pi)}(z,{\bf p}_{\perp}; {\bf s}) = \hat{f}_{(\alpha)}^{(q\rightarrow \pi)}(z,{\bf p}_{\perp}; {\bf s})
\nonumber \\
&+ \int {\cal D}^2 \eta \, \int {\cal D}^{4} p_{\perp} \,  \delta(z - \eta_1 \eta_2) \nonumber \\
&\times \delta^{(2)}({\bf p}_{\perp} - {\bf p}_{2 \perp} - \eta_2 {\bf p}_{1 \perp})
\hat{f}_{(\alpha)}^{(q \rightarrow Q)}(\eta_1, {\bf p}_{1 \perp} ; {\bf s}) \nonumber \\
&\times F_{(\alpha)}^{(Q \rightarrow \pi)} (\eta_2, {\bf p}_{2 \perp}; \langle {\bf S}_1 \rangle ) \,.   
\label{recurs2}
\end{align}
From this equation it follows that the ``favored'' combination 
${\displaystyle \frac{1}{3} F_{(0)}^{(q \rightarrow \pi)} + \frac{1}{2} F_{(1)}^{(q \rightarrow \pi)}}$ and
the ``neutral'' function $\frac{1}{3} F_{(0)}^{(q \rightarrow \pi)}$ 
have non-zero driving terms, while the ``unfavored'' combination 
${\displaystyle \frac{1}{3} F_{(0)}^{(q \rightarrow \pi)} - \frac{1}{2} F_{(1)}^{(q \rightarrow \pi)}}$
has no driving term, which is a simple consequence of charge conservation.

Second, we note that the momentum and isospin sum rules for the elementary FFs follow from
the general forms (\ref{moml}) - (\ref{isospin}), if the sum over $h$ includes both the
produced pion and the outgoing quark. Namely, the elementary counterpart of the LM sum rule
(\ref{moml}) is
\begin{eqnarray}
& & \hspace{-1cm} \int_0^1 {\rm d}z \, z \, \int{\rm d}^2 p_{\perp}  \left(
\sum_{\tau_{\pi}} \hat{d}^{(q \rightarrow \pi)}(z, {\bf p}_{\perp}^2)   \right.
\nonumber \\ 
& & + \left. 2 \sum_{\tau_{Q}} \hat{d}^{(q \rightarrow Q)}(z, {\bf p}_{\perp}^2) \right)
= 1\,,
\label{moms1}
\end{eqnarray}
that of the TM sum rule (\ref{momt}) is
\begin{eqnarray}
& & \hspace{-1cm} \int_0^1 \frac{{\rm d} z}{z} \, \int {\rm d}^2 p_{\perp} \, {\bf p}_{\perp}^2 \, 
\left( \frac{1}{m_{\pi}} \, \sum_{\tau_{\pi}} \, \hat{h}^{\perp (q \rightarrow \pi)}(z, {\bf p}_{\perp}^2) 
\right. \nonumber \\ 
& & + \left.\frac{2}{M} \,  \sum_{\tau_{Q}} \, \hat{h}^{\perp (q \rightarrow Q)}(z, {\bf p}_{\perp}^2) \right)
= 0 \,,  
\label{stsr1}
\end{eqnarray}
and that of the isospin sum rule (\ref{isospin}) is
\begin{eqnarray}
& & \hspace{-1cm} \int_0^1 {\rm d}z  \, \int{\rm d}^2 p_{\perp}  \left(
\sum_{\tau_{\pi}} \tau_{\pi} \, \hat{d}^{(q \rightarrow \pi)}(z, {\bf p}_{\perp}^2) \right.
\nonumber \\ 
& & + \left. 2 \sum_{\tau_{Q}} \frac{\tau_Q}{2} \hat{d}^{(q \rightarrow Q)}(z, {\bf p}_{\perp}^2) \right)
= \frac{\tau_q}{2} \,. 
\label{iso}
\end{eqnarray}
The sum rules (\ref{moms1}) - (\ref{iso}) just express the momentum and isospin conservation
laws for the elementary fragmentation process, and are therefore model independent. (Explicit
model forms for pseudoscalar (ps) and pseudovector (pv) pion-quark coupling are collected
in Appendix C.) We stress again that in the ``full'' sum rules (\ref{moml}) - (\ref{isospin}) the
summation $\Sigma_h$ refers only to the pions, because after an infinite chain of elementary
fragmentation processes the final quark remainder will have zero LM and, on average, also zero
TM and zero isospin $z$ component. We will confirm this point in the TMD jet model 
in the next Subsection and in Appendix E.

\subsection{Explicit forms of TMD jet integral equations and sum rules}

In this Subsection we give the explicit forms of the integral equations for the
spin independent ($D^{(q \rightarrow \pi)}$) and quark - spin dependent ($H^{\perp (q \rightarrow \pi)}$)
FFs and confirm the associated sum rules. For this, we have to insert the
elementary FFs for an incoming polarized quark and outgoing pion or unpolarized quark, as given by
(\ref{qqunpol}) and (\ref{relation}), into the integral equation (\ref{recurs}), and use the 
following expression for the
mean polarization density of the quark produced in the first step (see Eqs.(\ref{s1av}) and (\ref{qqunpol1}),
(\ref{sav1})):
\begin{align}
&<{\bf S}_1> = \frac{2}{\hat{f}^{(q \rightarrow Q)}(\eta_1, {\bf p}_{1 \perp}; {\bf s})} \nonumber \\ 
&\times \left[  \frac{1}{M \eta_1} {\bf p}'_{1 \perp} \, \, \hat{d}_T^{\perp (q\rightarrow Q)}(\eta_1, {\bf p}_{1 \perp}^2) 
+ {\bf s}_T \hat{h}_T^{(q \rightarrow Q)}(\eta_1, {\bf p}_{1 \perp}^2)   \right. \nonumber \\ 
& \left. + \frac{1}{M^2 \eta_1^2} \, 
{\bf p}_{1 \perp} \left({\bf s}_T \cdot {\bf p}_{1 \perp} \right) 
\hat{h}_T^{\perp (q \rightarrow Q)}(\eta_1, {\bf p}_{1 \perp}^2) \right] \,.   
\label{sav}
\end{align}
We then obtain for the product on the r.h.s. of (\ref{recurs}):
\begin{align}
&\hat{f}^{(q \rightarrow Q)} (\eta_1, {\bf p}_{1 \perp}; {\bf s})
\, F^{(Q \rightarrow \pi)}(\eta_2, {\bf p}_{2 \perp}; \langle {\bf S}_1 \rangle ) \nonumber \\
&= \hat{f}^{(q \rightarrow Q)} (\eta_1, {\bf p}_{1 \perp}; {\bf s}) 
\, D^{(Q \rightarrow \pi)} (\eta_2, {\bf p}_{2 \perp}^2)   \nonumber  \\
&+ \frac{2}{m_{\pi} \eta_2} \left[  \frac{1}{M \eta_1}  \left({\bf p}_{1 \perp} \cdot {\bf p}_{2 \perp} \right)
\, \hat{d}_T^{\perp (q\rightarrow Q)}(\eta_1, {\bf p}_{1 \perp}^2)  \right. \nonumber \\
& \left.+  
\left({\bf p}_{2 \perp} \times {\bf s}_T \right)^3 \, \hat{h}_T^{\perp (q \rightarrow Q)}(\eta_1, {\bf p}_{1 \perp}^2)
\right. \nonumber \\
&- \left. \frac{1}{M^2 \eta_1^2} \left({\bf p}_{1 \perp} \times {\bf p}_{2 \perp}\right)^3 \left({\bf s}_T \cdot 
{\bf p}_{1 \perp} \right)  \hat{h}_T^{\perp (q \rightarrow Q)}(\eta_1, {\bf p}_{1 \perp}^2) \right]
\nonumber \\
& \times  H^{\perp (Q \rightarrow \pi)} (\eta_2, {\bf p}_{2 \perp}^2) \, .
\label{prod}
\end{align}
Inserting everything into (\ref{recurs}) we obtain the following two coupled integral equations 
\footnote{Because the isoscalar and isovector integral equations have completely the same form
(see (\ref{recurs2})), we will
omit the isospin index $(\alpha)$ in some of the following equations for simplicity.}:
\begin{align}
&D^{(q \rightarrow \pi)}(z, {\bf p}_{\perp}^2) =  \hat{d}^{(q \rightarrow \pi)}(z, {\bf p}_{\perp}^2) \nonumber \\ 
&+ 2 \int {\cal D}^2 \eta \, \int {\cal D}^{4} p_{\perp} \,  
\delta(z - \eta_1 \eta_2) \delta^{(2)}({\bf p}_{\perp} - {\bf p}_{2 \perp} - \eta_2 {\bf p}_{1 \perp})
\nonumber \\
&\times \left[\hat{d}^{(q \rightarrow Q)}(\eta_1, {\bf p}_{1 \perp}^2) \,D^{(Q \rightarrow \pi)}(\eta_2, {\bf p}_{2 \perp}^2)  
+ \frac{1}{M m_{\pi} z} \right.   \nonumber \\
&\times \left. \left({\bf p}_{1 \perp} \cdot {\bf p}_{2 \perp}\right) \,  
\hat{d}_T^{\perp (q \rightarrow Q)}(\eta_1, {\bf p}_{1 \perp}^2) \,
H^{\perp (Q \rightarrow \pi)}(\eta_2, {\bf p}_{2 \perp}^2) \right] \,,
\label{si1}
\end{align}
\begin{align}
&\left({\bf p}_{\perp} \times {\bf s}_T\right)^3 H^{\perp (q \rightarrow \pi)}(z, {\bf p}_{\perp}^2)
= \left({\bf p}_{\perp} \times {\bf s}_T\right)^3 \hat{h}^{\perp (q \rightarrow \pi)}(z, {\bf p}_{\perp}^2) \nonumber \\
&+ 2 \int {\cal D}^2 \eta \, \int {\cal D}^{4} p_{\perp} \,  
\delta(z - \eta_1 \eta_2) \delta^{(2)}({\bf p}_{\perp} - {\bf p}_{2 \perp} - \eta_2 {\bf p}_{1 \perp})
\nonumber \\
&\!\times \! \left[ \! \frac{m_{\pi}}{M} \, \eta_2  \left({\bf p}_{1 \perp} \times {\bf s}_T \right)^3 
\hat{h}^{\perp (q \rightarrow Q)}(\eta_1, {\bf p}_{1 \perp}^2) 
D^{ (Q \rightarrow \pi)}(\eta_2, {\bf p}_{2 \perp}^2) \right.   \nonumber \\
& \left. + \left(\eta_1 \left({\bf p}_{2 \perp} \times {\bf s}_T\right)^3 \, 
\hat{h}_T^{(q \rightarrow Q)}(\eta_1, {\bf p}_{1 \perp}^2) - \frac{1}{M^2 \eta_1} 
\left({\bf s}_T \cdot {\bf p}_{1 \perp} \right)  \right. \right. 
\nonumber \\
& \left. \left. \times  \,
\left({\bf p}_{1 \perp} \times {\bf p}_{2 \perp}\right)^3 \,
\hat{h}_T^{\perp (q \rightarrow Q)}(\eta_1, {\bf p}_{1 \perp}^2) \right) \,  
H^{\perp (Q \rightarrow \pi)}(\eta_2, {\bf p}_{2 \perp}^2)   \right]   \,.
\label{sd1}
\end{align} 
At this stage, it is easy to confirm our previous comment about the vanishing contribution from the
last term $(\propto s_L)$ in the elementary version of (\ref{ftot}) for the $q \rightarrow Q$ case: 
Although this term contributes 
to (\ref{sav}) and (\ref{prod}), it vanishes in the integral equation (\ref{sd1}). Therefore
only the transverse quark polarization contributes to inclusive pion production. 

In order to obtain the integral equation for the function $H^{\perp (q \rightarrow \pi)}$
from (\ref{sd1}), it is necessary to use the delta function to integrate over ${\bf p}_{2\perp}$.
Using simple identities which follow from rotational invariance in the
transverse plane, we obtain 
\begin{align}
&H^{\perp (q \rightarrow \pi)}(z, {\bf p}_{\perp}^2)   \nonumber \\
&= \hat{h}^{\perp (q \rightarrow \pi)}(z, {\bf p}_{\perp}^2) 
+ 2 \int {\cal D}^2 \eta \,\delta(z-\eta_1 \eta_2) \int{\rm d}^2 p_{1 \perp} 
\nonumber \\
&\times \left[  \frac{m_{\pi}}{M} \, \eta_2  \, X 
\,\hat{h}^{\perp (q \rightarrow Q)}\left(\eta_1, {\bf p}_{1 \perp}^2\right) \,
D^{ (Q \rightarrow \pi)}(\eta_2, {\bf p}_{2 \perp}^2)   \right.   \nonumber \\ 
& \left. + \left(
\left(\eta_1 - z \, X \right) \hat{h}_T^{(q \rightarrow Q)}(\eta_1, {\bf p}_{1 \perp}^2) \,
+ \frac{1}{M^2 \eta_1}  
\left({\bf p}_{1 \perp}^2 - {\bf p}_{\perp}^2 \, X^2  \right)
\right. \right. \nonumber \\ 
&\left. \left. \times   \, 
\hat{h}_T^{\perp (q \rightarrow Q)}(\eta_1, {\bf p}_{1 \perp}^2) \right) \,   
H^{\perp (Q \rightarrow \pi)}(\eta_2, {\bf p}_{2 \perp}^2) \right] \,, 
\label{sd3}
\end{align}
where we denoted ${\displaystyle X \equiv \frac{{\bf p}_{\perp} \cdot {\bf p}_{1 \perp}}{{\bf p}_{\perp}^2}}$, and
${\bf p}_{2 \perp}^2 \equiv \left({\bf p}_{\perp} - \eta_2 {\bf p}_{1 \perp}\right)^2$. 
The two coupled integral equations (\ref{si1}) and (\ref{sd3}) constitute important results of our
investigation. 

We now wish to show that the momentum and isospin sum rules (\ref{moml})-(\ref{isospin}) are
valid in this TMD jet model. In the subsequent discussions, we will use the following notation for the
$n$th moment of any TMD function $A(z,{\bf p}_{\perp}^2)$
\footnote{We only need the cases $n=0$, where $A^{[0]}(z) = A(z)$ is the
integrated function, and $n=1$. Note that, with this naive definition,
the dimension of the $n=1$ moment is different from the $n=0$ case.}    
\begin{align}
A^{[n]}(z) = \int\, {\rm d}^2 p_{\perp} \,  \left({\bf p}_{\perp}^2\right)^n  
A(z,{\bf p}_{\perp}^2) \,,
\label{defm}
\end{align}
and adopt the notations
\begin{align}
\langle A(z) \rangle &= \int_0^1 {\rm d}z \, A(z) \,,  \nonumber \\
\left( A(\eta_1) \right) \otimes \left( B(\eta_2) \right)(z)  
&= \! \int  {\cal D}^2 \eta \, \delta(z-\eta_1 \eta_2)  A(\eta_1)  B(\eta_2)\,. 
\label{not}
\end{align}

First, the well known LM and the isospin sum rules follow immediately from (\ref{si1}): 
Integrating over
${\bf p}_{\perp}$, the second term in $[ \dots ]$ vanishes, which leaves us with the usual one-dimensional
convolution integral for the spin independent FF\cite{IBCTY09}. For the isoscalar case we obtain
the LM sum rule
\begin{align}
&\langle z \, D_{(0)}^{(q \rightarrow \pi)}(z) \rangle \nonumber \\
&= \langle z \, \hat{d}_{(0)}^{(q \rightarrow \pi)}(z) \rangle 
+ 2 \langle z \,\hat{d}_{(0)}^{(q \rightarrow Q)}(z) \rangle \, \langle z D_{(0)}^{(Q \rightarrow \pi)}(z) \rangle
\nonumber \\
&= \langle z \, \hat{d}_{(0)}^{(q \rightarrow \pi)}(z) \rangle +  \langle (1-z) \, \hat{d}_{(0)}^{(q \rightarrow \pi)}(z) \rangle \,
\langle z \, D_{(0)}^{(q \rightarrow \pi)}(z) \rangle \,, 
\label{es} 
\end{align}
where we performed the shift $z \rightarrow (1-z)$ of the
integration variable. If we write (\ref{es}) formally as $R = r + r' R$, then $r+r'=1$ due to the
normalization (\ref{fnorm}), and we get $R=1$, as in the original quark jet model\cite{FF78}:
\begin{align}
\int_0^1 {\rm d}z \, z \, \int {\rm d}^2 p_{\perp} \, D_{(0)}^{(q \rightarrow \pi)}(z, {\bf p}_{\perp}^2)
= 1 \,,
\label{long}
\end{align}
which is Eq. (\ref{moml}) for the present case of $h = \pi$ only.  
For the isospin sum rule, we can simply use the model independent normalizations of the isovector
splitting functions listed in Appendix C to obtain
\begin{align}
\langle D_{(1)}^{(q\rightarrow \pi)}(z) \rangle &= \langle \hat{d}_{(1)}^{(q\rightarrow \pi)}(z) \rangle
+ 2 \langle \hat{d}_{(1)}^{(q\rightarrow Q)}(z) \rangle \langle D_{(1)}^{(Q \rightarrow \pi)} \rangle
\nonumber \\ 
&= \frac{2}{3} - \frac{1}{3} \langle D_{(1)}^{(q \rightarrow \pi)} \rangle \,.
\label{ev}
\end{align}
From this we obtain the isospin sum rule
\begin{align}
\int_0^1 {\rm d}z \, \int {\rm d}^2 p_{\perp} \, D_{(1)}^{(q \rightarrow \pi)}(z, {\bf p}_{\perp}^2)
= \frac{1}{2} \,,
\label{isospin1}
\end{align}
in agreement with (\ref{isospin}). 

Second, in order to confirm also the validity of the TM sum rule, we first
derive the integral equation for the $n=1$ moment $H^{\perp [1] (q \rightarrow \pi)}(z)$. 
For this, we multiply (\ref{sd3}) by ${\bf p}_{\perp}^2$, integrate and perform the shift ${\bf p}_{\perp}
\rightarrow {\bf p}_{\perp} + \eta_2 {\bf p}_{1 \perp}$. Using simple identities which follow from 
rotational invariance 
in the transverse plane, and expressing $z= \eta_1 \eta_2$ everywhere, we obtain the 
following simple one-dimensional integral equation:
\begin{eqnarray}
& &H^{\perp[1](q \rightarrow \pi)}(z) = \hat{h}^{\perp[1](q \rightarrow \pi)}(z)  \nonumber \\
& & + 2 \frac{m_{\pi}}{M}  \, \left(\hat{h}^{\perp [1] (q \rightarrow Q)}(\eta_1)\right) \,
\otimes \left( \eta_2^2 \, D^{(q \rightarrow \pi)}(\eta_2)\right)
\nonumber \\ 
& & + 2 \left(\eta_1 \, \hat{h}^{(q \rightarrow Q)}(\eta_1) \right) 
\otimes \, \left(H^{{\perp}[1](q \rightarrow \pi)}(\eta_2) \right) \,,
\label{oneint} 
\end{eqnarray}
where we defined the function
\begin{eqnarray}
& & \hat{h}^{(q \rightarrow Q)}(\eta) = \hat{h}_T^{(q \rightarrow Q)}(\eta) + \frac{1}{2 M^2 \eta^2}
\hat{h}_T^{\perp [1] (q \rightarrow Q)}(\eta) \,.   \nonumber \\
\label{hh}
\end{eqnarray}
For the sum rule (\ref{momt}) we need to divide (\ref{oneint}) by $2 z m_{\pi}$, which gives
\begin{eqnarray}
& &\frac{1}{2 z m_{\pi}}\, H^{\perp [1] (q \rightarrow \pi)}(z) =
\frac{1}{2 z m_{\pi}}\, \hat{h}^{\perp [1] (q \rightarrow \pi)}(z) \nonumber \\
& &+  \left(\frac{1}{M \eta_1} \, \hat{h}^{\perp [1] (q \rightarrow Q)}(\eta_1)\right) \otimes \left(
\eta_2 \,\,D^{(q \rightarrow \pi)}(\eta_2)\right)   \nonumber \\
& &+ 2 \, \left(\hat{h}^{(q \rightarrow Q)}(\eta_1) \right) \otimes \left( \frac{1}{2 \eta_2 m_{\pi}} \,  
H^{\perp [1] (q \rightarrow \pi)}(\eta_2) \right) \,.  \nonumber \\
\label{oneint1}
\end{eqnarray}
If we integrate Eq.(\ref{oneint1}) for the isoscalar parts over $z$ and use the 
LM sum rule (\ref{long}) and the relation (\ref{stsr1}) for the
elementary splitting functions, we see that the first two terms on the r.h.s. of (\ref{oneint1}) 
cancel each other in the integral. What remains is the following relation:
\begin{align}
\int_0^1 \frac{{\rm d}z}{2 z m_{\pi}} \,  H_{(0)}^{\perp [1] (q \rightarrow \pi)}(z) = 
C \times \int_0^1 \frac{{\rm d}z}{2 z m_{\pi}} \,  H_{(0)}^{\perp [1] (q \rightarrow \pi)}(z) \,,
\label{t}
\end{align}
where we defined the constant
\begin{eqnarray}
& & C = 2 \int_0^1 {\rm d}z \hat{h}_{(0)}^{(q \rightarrow Q)}(z)      
\label{y}  \\
& & = \left( \int_0^1 {\rm d} z \, h_{(0)}^{(q \rightarrow Q)}(z) \right) \cdot
\left( \int_0^1 {\rm d} z \, d_{(0)}^{(q \rightarrow Q)}(z) \right)^{-1} \,,
\nonumber \\
\label{y1}
\end{eqnarray}
where in the second step we used $\hat{h}^{(q\rightarrow Q)}(z) = h^{(q\rightarrow Q)}(z)/(1-Z_Q)$ 
with $(1-Z_Q)$ from (\ref{zform}).
From (\ref{t}) we see that, unless $C = 1$, the isoscalar TM sum rule must 
vanish.
On general grounds, $|C| \leq 1$ follows from one of the positivity bounds for
the twist-2 quark FFs: Because the $q \rightarrow Q$ FF has the physical 
interpretation of the distribution function of $Q$ inside $q$ (see Sect. II), 
we see that $h^{(q \rightarrow Q)}$ is the transversity distribution function 
and $d^{(q \rightarrow Q)}$ the unpolarized distribution function of $Q$ inside $q$.
The probabilistic interpretation of those functions leads to the positivity
bound 
$|h^{(q\rightarrow Q)}(z)| \leq d^{(q \rightarrow Q)}(z)$ \cite{JJ92},
which can be extended \cite{BBHM00} to the TM dependent functions:
$|h^{(q\rightarrow Q)}(z, {\bf p}_{\perp}^2)| \leq
d^{(q \rightarrow Q)}(z,{\bf p}_{\perp}^2)$. This inequality immediately leads to
$|C| \leq 1$. The boundary value $C=1$ would correspond to the case where
$h^{(q\rightarrow Q)}$ and $d^{(q \rightarrow Q)}$ are identical functions of $z$ and 
${\bf p}_{\perp}^2$, which we exclude here
\footnote{Writing $h^{(q\rightarrow Q)} = f_{\uparrow} - f_{\downarrow}$ and
$d^{(q \rightarrow Q)} = f_{\uparrow} + f_{\downarrow}$ with semi-positive definite
functions $f_{\uparrow}$ and $f_{\downarrow}$, the
boundary value 
$C=1$ would mean that $f_{\downarrow} = 0$, i.e; the probability 
distribution of quarks
with transversity opposite to the parent quark would have to vanish identically for
all values of $z$ and ${\bf p}_{\perp}^2$.}.
Actually, for the case of pion emission, the result for $C$ 
obtained for both ps and pv quark-pion coupling
shows that $-1<C<0$ (see Appendix C). \\

Finally in this Section, we add the following three comments:
\begin{itemize}
\item 
In our present TMD jet model, the constant $C$ of (\ref{y}) 
gives the ratio of the 
mean polarizations of the outgoing and incoming quarks (including a sum over the outgoing quark
flavors) for one elementary fragmentation step, i.e., a measure
for the quark depolarization. Taking the first step as an example, 
this follows from the form given by (\ref{sav1}):
\begin{eqnarray}
& & \int_0^1 {\rm d}\eta \int {\rm d}^2 p_{\perp} \, \sum_{\tau_Q} \hat{{\bf f}}^{(q\rightarrow Q)}
(\eta, {\bf p}_{\perp}; {\bf s}) = C \, {\bf s}_T \,. \nonumber \\
\label{depol}
\end{eqnarray}
\item 
The finite constituent quark mass $M$ causes mixing of operators with opposite 
chirality in the integral equation (\ref{si1}): We remind that the Dirac matrices $\gamma^+$ and
$\gamma^+ \gamma_5$ of (\ref{vplus}) and (\ref{aplus}) are chiral even (anticommute
with $\gamma_5$), while $i \sigma^{i+} \gamma_5$ of (\ref{tplusi}) is chiral odd (commutes
with $\gamma_5$). If there were no mass term in the quark propagator, operators with
opposite chirality could not couple in the integral equation.
Therefore the term $\propto \hat{d}_T^{\perp (q \rightarrow Q)} \,
H^{\perp (Q \rightarrow \pi)}$ in the integral equation (\ref{si1}) arises entirely from
the finite constituent quark mass term in the propagators. (Explicit model examples 
to illustrate this point are discussed in Appendix C for both ps and pv
pion-quark coupling.)
\item
The integral equations derived in this Section and the associated sum rules hold in any 
effective quark theory which does not involve explicit gluon and gauge link degrees of freedom,
and which satisfies the following 3 points which were used in the verification of the
TM sum rule in the steps from Eq.(\ref{oneint}) to (\ref{y1}): 
(i) the LM sum rule (\ref{long}), (ii) the TM conservation in each fragmentation step expressed by 
(\ref{stsr1}), and (iii) the quark depolarization factor $C$ 
of (\ref{y}) is not equal to unity, i.e., the transversity distribution function and
the unpolarized distribution function of a quark inside a parent quark are not identical
to each other.
\end{itemize}   

\section{Summary}

The analysis of TMD quark distribution and fragmentation functions is a very active
field of present experimental and theoretical research. For the description of quark 
TMD distribution functions, one can follow the methods based on relativistic bound state
vertex functions for hadrons, which have been applied successfully to form factors
and the longitudinal quark momentum distributions. For the description of quark FFs, 
however, one has to consider multi-fragmentation processes, where the quark
produces a cascade of mesons. One purpose of this paper was therefore to formulate the
TMD jet model, which is suitable for numerical calculations in effective quark theories.
Limiting ourselves to the case of inclusive pion production for simplicity and clarity, 
we used a product ansatz for the TMD FF, similar to that used by Field and 
Feynman for the description of longitudinal quark jets\cite{FF78}.
From this product ansatz we derived the integral equations for the spin independent
and quark - spin dependent FFs. The proper treatment of the spin of the
quarks in the intermediate states requires the use of several elementary TMD splitting functions
in the integral equations. We found that these integral
equations are coupled to each other, that is, the spin independent and quark - spin dependent FFs
are mutually interrelated.
We showed that in this TMD jet model all momentum and isospin sum rules are satisfied. 
This is possible because after many hadron emissions the final quark remainder has
zero longitudinal momentum and, on average, also zero transverse 
momentum and zero $z$- component of isospin.   

The numerical solutions of the integral equations derived in this paper, using model
input splitting functions, will allow to obtain the relevant FFs in future work.
An important task thereby will be to extend the framework to additional hadron production
channels, such as kaons, vector mesons and their strong decays, as well as baryons. 
The Monte-Carlo method will be naturally suited for
this purpose, which can also allow to study various correlations between FFs describing
single - and multi-hadron inclusive production. 
In order to make contact to experiment, it is also important to take into account 
the $Q^2$ evolution of the calculated TMD FFs\cite{APR12}. Together with the model TMD 
PDFs, they can be used to calculate observables like cross sections and asymmetries
for various SIDIS processes. Finally,
in view of recent experimental analyses\cite{ATLAS15}, it is of great interest to 
explore quark FFs in the nuclear medium.

\vspace{1 cm}

\noindent
{\sc Acknowledgments}

This work was supported by the Grant in Aid (Kakenhi) of the Japanese Ministry of
Education, Culture, Sports, Science and Technology, Project No. 25400270; the
Australian Research Council through the ARC Centre of Excellence for Particle
Physics at the Terascale (CE110001104); and an ARC Australian Laureate
Fellowship FL0992247 and Discovery Project DP151103101.

\appendix

\section{Transverse Lorentz transformations}
\setcounter{equation}{0}
A transverse Lorentz transformation is defined so as to leave the component 
$a^+ = a_-$ of any 4-vector $a_{\mu} = \left(a_+, a_-, a_1, a_2 \right)$ unchanged.
It involves the parameters $b_-$ and ${\bold b}_T$, and the Lorentz matrix is
expressed by\cite{MD01}
\begin{align}
\Lambda_{\mu}^{\,\,\,\nu} = \left(
\begin{array}{cccc}
1 & \frac{{\bold b}_T^2}{2 b_-^2} & \frac{b^1}{b_-} & \frac{b^2}{b_-} \\
0 & 1 & 0 & 0 \\
0 & - \frac{b_1}{b_-} & 1 & 0 \\
0 & - \frac{b_2}{b_-} & 0 & 1 
\end{array}
\right) \,.
\label{lmat}
\end{align}
The quark and hadron momenta are transformed as
$k_{\mu}' = \Lambda_{\mu}^{\,\,\,\nu} \, k_{\nu} \,$,
$\,p_{\mu}' = \Lambda_{\mu}^{\,\,\,\nu} \, p_{\nu}$.
If we start from a system S, where in general both ${\bf p}_T$ and ${\bf k}_T$
are nonzero, we consider the following two cases:
(1) By using $b_-=k_-$, ${\bold b}_T = {\bold k}_T$ in (\ref{lmat}), we arrive at a system 
S' where ${\bold k}'_T = 0$. The relation between the transverse momenta in this case becomes
${\bold p}_T' = {\bold p}_T - z {\bold k}_T$. 
(2) By using $b_-=p_-$, ${\bold b}_T = {\bold p}_T$ in (\ref{lmat}), we arrive at a system 
S' where ${\bold p}'_T = 0$. The relation between the transverse momenta in this case becomes 
${\bold k}_T' = {\bold k}_T - \frac{{\bold p}_T}{z}$.

We note that one can express the above transformations also in usual Minkowski
coordinates. For example, for the transformation (1) discussed above we get
\begin{eqnarray}
p_0' &=& p_0 + \frac{1}{2 \sqrt{2} p_-} \left({\bf k}_T^2 z^2 - 2 z {\bf k}_T \cdot
{\bf p}_T \right) \,,\label{p0} \\
p_3' &=& p_3 + \frac{1}{2 \sqrt{2} p_-} \left({\bf k}_T^2 z^2 - 2 z {\bf k}_T \cdot
{\bf p}_T \right) \,, \label{p3}
\end{eqnarray}
and one can confirm that $p_0^{'2} - p_3^{'2} - {\bf p}_T^{'2} =p_0^2 - p_3^2 - {\bf p}_T^2$.
Therefore, at leading order (leading power of $p_-$), the
direction ${\hat {\bf p}}$ is always in the 3 - direction, and the corrections to this
are of subleading order.

The operation used in the definition of the quark decay matrix (see Eq.(\ref{ndef}))
\begin{align}
\frac{1}{2 z} \int \frac{{\rm d}k_+ \, {\rm d}k_-}{(2\pi)^4} 
\delta \left( \frac{1}{z} - \frac{k_-}{p_-} \right) 
\label{op}
\end{align}
is invariant under the transverse Lorentz transformations, because the transformation of
$k_+$ can be eliminated by a shift of the integration variable. We also note that the  
vectors ${\bold s}$ and ${\bold S}$ in the parametrization of all FFs used in this
paper (see Eq.(\ref{ftot})) are not subject
to any Lorentz transformation, because by definition they denote generic (constant) 
vectors in space; i.e., parameters which specify the spin 4-vector (see for example Eq.(\ref{spin4})). 
Quantities like $S_L$, ${\bf S}_T$, for example, are defined by 
$S_L = \left({\bf S} \cdot {\hat {\bf {\bf p}}}\right)$
and ${\bf S}_T = {\bf S} - {\hat {\bf p}} \left({\bf S} \cdot {\hat {\bf p}}\right)$, and 
for the leading produced particle (leading twist) the direction ${\hat {\bf p}}$ is not changed under the 
transverse Lorentz transformation as discussed above. 
We therefore arrive at the following simple rule
for the transverse Lorentz transformation of any FF:
\begin{align}
F(z,{\bf p}_{T}, {\bf k}_T; {\bf S}, {\bf s})
=  F(z, {\bf p}_T - z {\bf k}_T; {\bf S}, {\bf s} | {\bf k}_{\perp}=0) \,.
\label{general}
\end{align}
Here the notation on the r.h.s. refers to a frame where the transverse momentum of the
fragmenting quark vanishes, and in this case the parametrization given by Eq.(\ref{ftot}) 
holds. Namely, in a general system, we simply have to replace 
the momentum ${\bf p}_{\perp}$ in Eq.(\ref{ftot}) according to
${\bf p}_{\perp} \rightarrow {\bf p}_T - z {\bf k}_T$.

\section{Light front spinors and Melosh rotation}
\setcounter{equation}{0}
The positive energy spinor in the usual Dirac representation is given by
\begin{align}
u_{\lambda}(p) = \sqrt{E + m} \left(
\begin{array}{c}
\hat{\chi}_\lambda  \\ 
\frac{\boldsymbol{\sigma}\cdot {\bf p}}{E+m} \, \hat{\chi}_\lambda 
\end{array}
\right) \,,
\label{spinor}
\end{align}
where $\hat{\chi}_\lambda$ is a 2-component Pauli spinor. In this Appendix we denote the mass by $m$, 
the energy $E_p$ by $E$, 
and the normalization is $\overline{u} u = 2m$.
The ``good component'' of the spinor is obtained from (\ref{spinor}) by applying the
projection operator Eq.(\ref{plus}):
\begin{align}
&u_{+ \lambda}(p) = \Lambda_{(+)} u_{\lambda} (p) = \frac{1}{2} \sqrt{E + m} \left(
\begin{array}{c} 
{\bf 1} + \frac{\sigma_3 \left(\boldsymbol{\sigma}\cdot {\bold p}\right)}{E + m} \\
\sigma_3 + \frac{\left(\boldsymbol{\sigma}\cdot {\bold p}\right)}{E + m} 
\end{array} \right) 
\hat{\chi}_\lambda  \label{up1} \\ 
&= \sqrt{\frac{E + p^3}{2}} \left( 
\begin{array}{c}
U_M^{\dagger}  \\
\sigma_3 U_M^{\dagger} 
\end{array}  \right)
\hat{\chi}_\lambda  
\equiv 
\sqrt{\frac{E + p^3}{2}} \left( 
\begin{array}{c}
{\bf 1} \\
\sigma_3 
\end{array}  \right) \,
\chi_\lambda \,. \nonumber \\
\label{up2}
\end{align}
Here the Pauli spinor $\chi_\lambda$ is defined by 
$\hat{\chi}_\lambda = U_M \, \chi_\lambda$,
and $U_M$ is the so called ``Melosh rotation''\cite{MEL74}. (The explicit form of
the spinor rotation $U_M$ can easily be obtained from the above relations.)

Using the form of the spinor $u_{+ \lambda}(p)$ given in (\ref{up2}), the relation
(\ref{spin}) of Sect. II can easily be shown as follows:
\begin{align}
&\overline{u}_{\lambda'}(p)\, \Gamma^{\mu} \, u_{\lambda}(p) =
\sqrt{2} \, u_{+ \lambda'}^{\dagger}(p) \, \tilde{\Gamma}^{\mu}\, u_{+ \lambda}(p)
\nonumber \\
&= \sqrt{2} \, \frac{E+p^3}{2}\, \chi^{\dagger}_{\lambda'} \left(1,\sigma_3\right) \tilde{\Gamma}^{\mu}
\left(
\begin{array}{c}
1 \\
\sigma_3 
\end{array}
\right) \chi_\lambda  \nonumber \\
&= \sqrt{2} \, (E+p^3) \, \chi^{\dagger}_{\lambda'} \sigma^{\mu} \chi_\lambda \,
= 2 p_- \, \left( \sigma^{\mu} \right)_{\lambda' \lambda} \,,
\label{pr}
\end{align}
where we used the definitions of $\Gamma^{\mu}$, Eq. (\ref{gamma}), and
$\Gamma^{\mu} = \gamma^+ \tilde{\Gamma}^{\mu}$. Eq.(\ref{pr}) is the same as (\ref{spin}) of
Sect. II.

The quantity (\ref{pr}) represents a Hermitian $2 \times 2$ matrix in the
spin indices $(\lambda', \lambda)$, and contraction with $s_{\mu} = (1, {\bf s})$
leads to  
$\rho_{\lambda' \lambda}({\bf s}) = 
\frac{1}{2} \left(1 + {\bf s} \cdot {\boldsymbol \sigma}\right)_{\lambda' \lambda}$ 
of (\ref{f3}) in the main text. Denoting by $s$ the magnitude of the 
polarization vector ($0 \leq s \leq 1$) and by ${\bf \hat{s}}$ its direction, 
the operator $\rho({\bf s}) = 
\frac{1}{2} \left(1 + {\bf s} \cdot {\boldsymbol \sigma}\right)$ 
can be written in the form
\begin{align}
\rho({\bf s}) = w_+ \, \frac{1}{2} \left(1 + {\bf \hat{s}} \cdot {\boldsymbol \sigma}\right)
+ w_- \, \frac{1}{2} \left(1 - {\bf \hat{s}} \cdot {\boldsymbol \sigma}\right) \,, 
\label{rhopm}
\end{align}
where $w_{\pm} = \frac{1}{2} \left(1 \pm s\right)$. 
Therefore, for a fully polarized quark ($s=1$),  
$\rho({\bf s})$ becomes a projector onto the direction ${\bf \hat{s}}$, while 
for a partially polarized quark ($s< 1$),
$\rho({\bf s})$ is a linear combination of the 
projectors onto the directions ${\bf \hat{s}}$ and $-{\bf \hat{s}}$ with coefficients
$w_+$ and $w_-$, respectively. Therefore $\rho({\bf s})$ 
can be identified with the usual spin density matrix.

For easier interpretation of some of the relations in the main text,
we finally give the form of the spin density matrix in the basis which
diagonalizes ${\bf s}\cdot {\boldsymbol \sigma}$:
\begin{align}
\rho_{\lambda' \lambda}({\bf s}) = \delta_{\lambda' \lambda} \frac{1}{2} \left(1 + s \, \lambda\right) \,,
\label{bas}
\end{align}
where $\lambda = \pm 1$.   
In this basis, the spin average of any quantity $A$ takes the form
\begin{align}
{\rm Tr} \left(\rho({\bf s}) \, A \right) = w_+ \, A_{11} + w_- \, A_{-1 -1}  \,.  
\label{tr}
\end{align}

\section{Explicit forms of elementary fragmentation functions}
\setcounter{equation}{0}

\begin{figure}
\begin{center}
\subfigure{\includegraphics[scale=0.6]{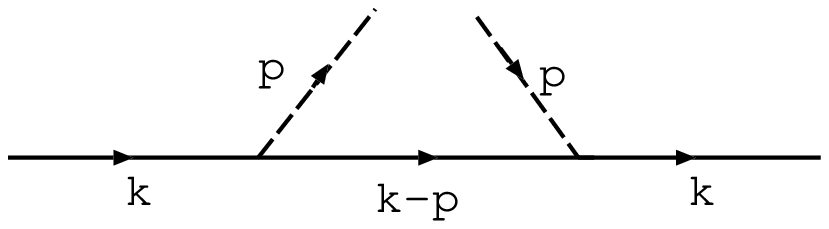}} \
\subfigure{\includegraphics[scale=0.6]{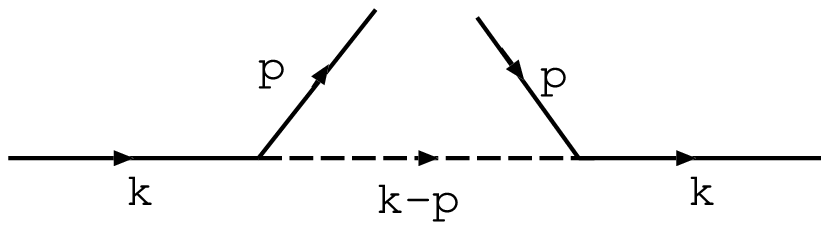}}
\caption{Cut diagrams for elementary fragmentation processes $q \rightarrow \pi$ (top) and $q \rightarrow Q$ (bottom).
The solid line denotes a quark, and the dashed line a pion. The cut goes through the line labeled by the
momentum $k-p$.}
\end{center}
\end{figure}

In this Appendix we list model results for the elementary $q \rightarrow \pi$ and $q \rightarrow Q$
splitting functions, parametrized as in Eq.(\ref{ftot}), and their 
sum rules. 
We will mainly refer to the case of ps coupling of constituent quarks
(mass $M$) to pions, but also discuss the results for pv coupling in
those cases which serve to illustrate the model independence of the points discussed 
in Sect. III of the main text.

The non-fragmentation term $\propto Z_Q$ of Eq.(\ref{fg}) is easily obtained 
from the operator definitions (\ref{ndef}) and (\ref{Ndef}) as the contribution of the
hadronic vacuum state $|0 \rangle$ to the sum $\Sigma_n$ in (\ref{Ndef}). 
All 4 operators (\ref{gamma}) contribute to this term, and by using (\ref{spin}) for the spinor
matrix elements one easily derives the spin dependence as expressed in (\ref{fg}).
Another way to see this is to use the formal analogue of Eq.(\ref{eend}) for the
``$0^{\rm th}$ step'': 
$f_0({\bf S}, {\bf s}) \equiv {\rm Tr} \left( \tilde{f}_0({\bf s}) \rho({\bf S}) \right)
= \frac{1}{2} \left(1  + {\bf S} \cdot {\bf s} \right)$, where 
$\tilde{f}_0({\bf s}) = \rho({\bf s})$ follows from setting $N=0$ in Eq.(\ref{form})
\footnote{The corresponding argument for pure spin states is to use the relation
$| {\bf S} \rangle \langle {\bf S}| = \frac{1}{2} \left(1 + {\bf S} \cdot {\boldsymbol \sigma}
\right)$, which implies $\langle {\bf s}|{\bf S} \rangle = \frac{1}{2} \left(1 + {\bf S} \cdot {\bf s}\right)$.}.
We do not list the non-fragmentation terms in the formulas of 
this Appendix, because eventually they can be absorbed
into the renormalized FFs, as explained in Appendix D.

The tree level cut diagrams of Fig.2 contribute to the
six $T$-even splitting functions of Eq.(\ref{ftot}), and in order to obtain non-zero results
for the $T$-odd functions $d_T^{\perp}$ and $h^{\perp}$ one has to consider the loop diagrams
shown in Figs.3 and 4\footnote{As as shown in Ref.\cite{BKMM01}, the other one-loop diagrams do not
contribute to the $T$-odd functions considered here.}. 
In order to facilitate comparison with previous works\cite{MMP10,ABM05,BGGM08}, we give the expressions for the case 
where a neutral pion is produced (for $q\rightarrow \pi$ case) or on the cut (for $q\rightarrow Q$ case), which
we refer to as the ``neutral functions''. 
The flavor dependence is then expressed in terms of those neutral functions by
\begin{align}
f^{(q \rightarrow \pi)} &= f_{\rm neutral}^{(q \rightarrow \pi)} \left(1 + \tau_q \tau_{\pi}\right) \,, \label{ff1} \\
f^{(q \rightarrow Q)} &= f_{\rm neutral}^{(q \rightarrow Q)} \left(\frac{3}{2} - \frac{\tau_q \tau_Q}{2}\right) \,.
\label{ff2}
\end{align}
Because of the definitions (\ref{isivqp}) and (\ref{isivqq}) of the main text,
the isoscalar and isovector functions can be obtained from the neutral ones by
\begin{align}
f_{(0)}^{(q \rightarrow \pi)} &= 3 f_{\rm neutral}^{(q \rightarrow \pi)} \,,\,\,\,\,\,\,\,\,\,\,\,\,
f_{(0)}^{(q \rightarrow Q)} = 3 f_{\rm neutral}^{(q \rightarrow Q)} \,,  \label{isos} \\
f_{(1)}^{(q \rightarrow \pi)} &= 2 f_{\rm neutral}^{(q \rightarrow \pi)} \,,\,\,\,\,\,\,\,\,\,\,\,\,
f_{(1)}^{(q \rightarrow Q)} = - f_{\rm neutral}^{(q \rightarrow Q)} \,. \label{isov} 
\end{align}

Consider first the tree diagrams of Fig.2 
for ps coupling. 
For the $q \rightarrow \pi$ fragmentation 
they give the well known result\cite{IBCTY09,MMP10}
\begin{align}
&d_{\rm neutral}^{(q \rightarrow \pi)}(z, {\bf p}_{\perp}^2)
= \frac{z}{2} \frac{g_{\pi}^2}{\left(2 \pi \right)^3} \frac{{\bf p}_{\perp}^2 + M^2 z^2}
{\left[{\bf p}_{\perp}^2 + M^2 z^2 + (1-z) m_{\pi}^2 \right]^2} \,, 
\label{old} 
\end{align}
while for the $q \rightarrow Q$ fragmentation they give the following six $T$-even functions\cite{MMP10}:
\begin{align}
&d_{\rm neutral}^{(q \rightarrow Q)}(z, {\bf p}_{\perp}^2) = 
\frac{1-z}{4}  \frac{g_{\pi}^2}{\left(2 \pi \right)^3}  
\frac{{\bf p}_{\perp}^2 + M^2 (1-z)^2}
{\left[{\bf p}_{\perp}^2 + M^2 (1-z)^2 + z m_{\pi}^2 \right]^2} \,,
\label{dqq}  \\
&h_{T,{\rm neutral}}^{(q \rightarrow Q)}(z, {\bf p}_{\perp}^2) = 
- d_{\rm neutral}^{(q \rightarrow Q)}(z, {\bf p}_{\perp}^2)  \,,
\label{htqq}  \\
&h_{T,{\rm neutral}}^{\perp (q \rightarrow Q)}(z, {\bf p}_{\perp}^2) = 
\frac{1-z}{2}  \frac{g_{\pi}^2}{\left(2 \pi \right)^3}  
\frac{M^2 z^2}
{\left[{\bf p}_{\perp}^2 + M^2 (1-z)^2 + z m_{\pi}^2 \right]^2} \,,
\label{htpqq} \\
&h_{L,{\rm neutral}}^{\perp (q \rightarrow Q)}(z, {\bf p}_{\perp}^2) = 
 \frac{1-z}{2}  \frac{g_{\pi}^2}{\left(2 \pi \right)^3}  
\frac{M^2 z (1-z)}
{\left[{\bf p}_{\perp}^2 + M^2 (1-z)^2 + z m_{\pi}^2 \right]^2} \,,
\label{hlpqq} \\
&g_{L,{\rm neutral}}^{(q \rightarrow Q)}(z, {\bf p}_{\perp}^2) = \frac{1-z}{4}  \frac{g_{\pi}^2}{\left(2 \pi \right)^3}  
\frac{-{\bf p}_{\perp}^2 + M^2 (1-z)^2}
{\left[{\bf p}_{\perp}^2 + M^2 (1-z)^2 + z m_{\pi}^2 \right]^2} 
\label{glqq}  \\
&g_{T,{\rm neutral}}^{(q \rightarrow Q)}(z, {\bf p}_{\perp}^2) = 
h_{L,{\rm neutral}}^{\perp (q \rightarrow Q)}(z, {\bf p}_{\perp}^2)  \,.
\label{gtqq} 
\end{align} 

\begin{figure}
\begin{center}
\includegraphics[scale=0.45]{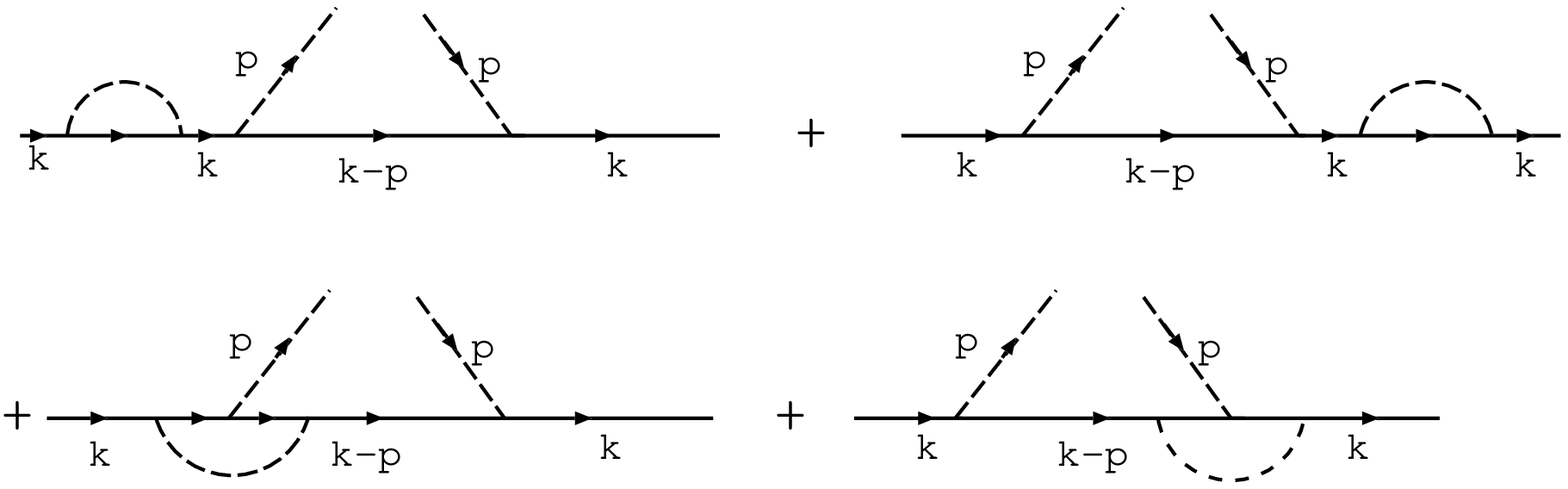}
\caption{Cut diagrams with a pion loop for the elementary fragmentation process $q \rightarrow \pi$.
The solid line denotes a quark, and the dashed line a pion. 
The cut goes through the line labeled by the momentum $k-p$.}
\end{center}
\end{figure}
 
\begin{figure}
\begin{center}
\includegraphics[scale=0.45]{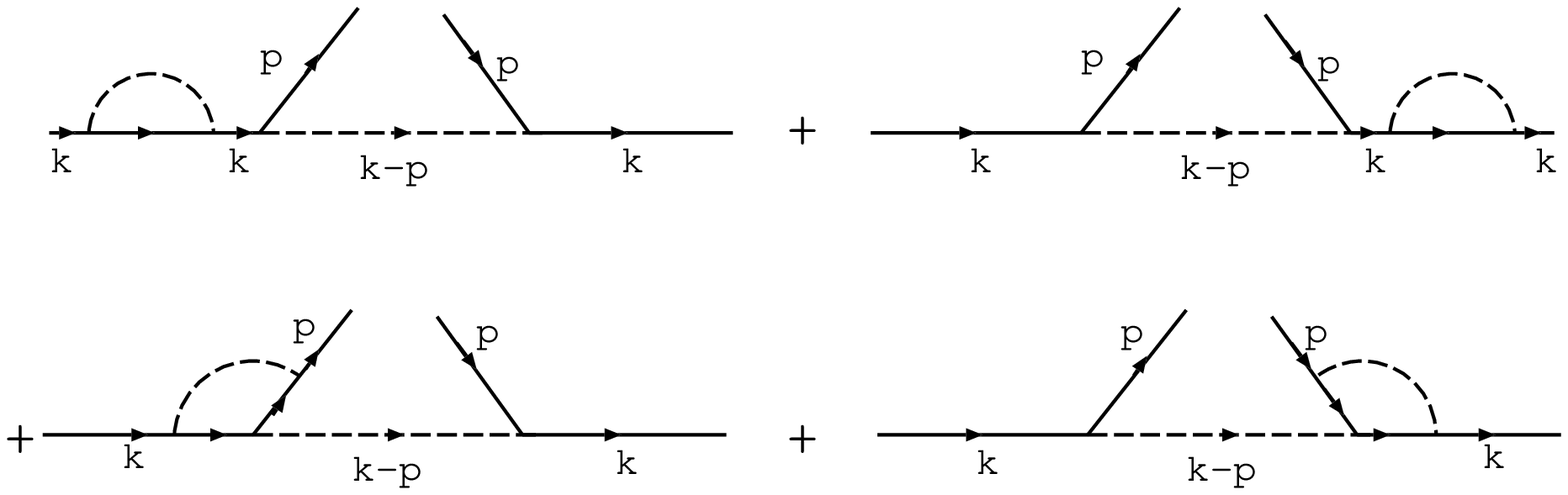}
\caption{Cut diagrams with a pion loop for the elementary fragmentation process $q \rightarrow Q$.
The solid line denotes a quark, and the dashed line a pion. 
The cut goes through the line labeled by the momentum $k-p$.}
\end{center}
\end{figure}

The pion loop diagrams of Fig.3 give the following results for the elementary $T$-odd
$q \rightarrow \pi$ FF 
for the case of ps coupling
\cite{ABM05,BGGM08}:
\begin{align}
&h_{\rm neutral}^{\perp (q \rightarrow \pi)}(z,{\bf p}_{\perp}^2)  \nonumber \\
&= - \frac{g_{\pi}^2}{(2 \pi)^3} \frac{M \, m_{\pi}}{1-z} 
\left( \frac{{\rm Im} \tilde{\Sigma}(k^2)}{(k^2 - M^2)^2}
+ \frac{{\rm Im} \tilde{\Gamma}_{\pi}(k^2)}{k^2 - M^2} \right) \,,
\label{collpi}  
\end{align}
where the whole expression should be taken at
\begin{align}
k^2 = \frac{1}{z(1-z)}\left({\bf p}_{\perp}^2 + M^2 z + m_{\pi}^2(1-z)\right) \,.
\nonumber
\end{align}
In (\ref{collpi}) we have $\tilde{\Sigma} = A + B$ and
$\tilde{\Gamma}_{\pi} = D + E + M\,F$, where the various functions are defined by the
representation of the quark self energy $\Sigma$ and the $qq\pi$ vertex correction $\Gamma_{\pi}$ 
in terms of
Dirac matrices as $\Sigma = A \fslash{k} + B\,M$ and 
$\Gamma_{\pi}(k,p) = C + D \,\fslash{p} + E \,\fslash{k} + F \,\fslash{p} \, \fslash{k}$. 
The analytic forms of ${\rm Im} \tilde{\Sigma}$ and ${\rm Im} \tilde{\Gamma}_{\pi}$
are given by\cite{ABM05} 
\begin{align}
&{\rm Im} \tilde{\Sigma}(k^2) = \frac{3 g_{\pi}^2}{16 \pi^2} \left(1 - \frac{M^2 - m_{\pi}^2}{k^2}\right)
I_1 \,, \label{sigmat} \\
&{\rm Im} \tilde{\Gamma}_{\pi}(k^2) = \frac{g_{\pi}^2}{8 \pi^2} \frac{k^2 - M^2 + m_{\pi}^2}{\lambda}
\nonumber \\
&\times \left(I_1 + \left(k^2 - M^2 - 2 m_{\pi}^2 \right) I_2 \right) \,,
\label{vertpit}
\end{align}
where the integrals $I_1$ and $I_2$ are given by
\begin{align}
&I_1 = \int {\rm d}^4 \ell \, \delta(\ell^2 - m_{\pi}^2) \, \delta\left[(k-\ell)^2 - M^2 \right]
\nonumber \\
&= \frac{\pi}{2 k^2} \sqrt{\lambda} \, \Theta(k^2 - (M+m_{\pi})^2)  \,,
\nonumber \\
&I_2 = \int {\rm d}^4 \ell \, \frac{\delta(\ell^2 - m_{\pi}^2) \, \delta\left[(k-\ell)^2 - M^2 \right]}
{(k-p-\ell)^2 - M^2}  \nonumber \\
&= - \frac{\pi}{2 \sqrt{\lambda}} \, {\rm log}\left(1+ \frac{\lambda}{k^2 M^2 - (M^2-m_{\pi}^2)^2}\right)
\nonumber \\
&\times \Theta(k^2 - (M+m_{\pi})^2)  \,,
\nonumber
\end{align}
and the function $\lambda$ is given by
\begin{align}
\lambda(k^2) = \left(k^2 - (M+m_{\pi})^2\right)  \left(k^2 - (M-m_{\pi})^2\right) \,. 
\nonumber
\end{align}

For the elementary $T$-odd
$q \rightarrow Q$ FFs, the pion loop diagrams of Fig. 4 give the following results
for ps coupling
\cite{MMP10}:
\begin{align}
&h_{\rm neutral}^{\perp (q \rightarrow Q)}(z,{\bf p}_{\perp}^2) = \nonumber \\
&= \frac{1}{2}
\frac{g_{\pi}^2}{(2 \pi)^3} \frac{M^2}{1-z} 
\left( \frac{ {\rm Im} \tilde{\Sigma}(k^2)}{(k^2 - M^2)^2}
+ \frac{{\rm Im} \tilde{\Gamma}_{q}(k^2)}{k^2 - M^2} \right) \,,
\label{collq} 
\end{align}
\begin{eqnarray}
d_{T}^{\perp (q \rightarrow Q)}(z,{\bf p}_{\perp}^2) &=& -h^{\perp (q \rightarrow Q)}(z,{\bf p}_{\perp}^2) \,,  
\label{dq}
\end{eqnarray}
where the expressions should be taken at
\begin{eqnarray}
k^2 = \frac{1}{z(1-z)}\left({\bf p}_{\perp}^2 + (1-z) M^2 + m_{\pi}^2 z \right)  \,.
\nonumber
\end{eqnarray}
For some fixed value of $k^2$ one has
${\rm Im} \tilde{\Gamma}_{q}(k^2) = {\rm Im} \tilde{\Gamma}_{\pi}(k^2)$.

The above model expressions illustrate some general features discussed in the main text.
First, the validity of the TM sum rule for the elementary FFs is evident from Eqs.(\ref{collpi}) and
(\ref{collq}). Second, if we insert the above model expressions into the expression
(\ref{y1}) for the quark depolarization factor $C$ we obtain  
%
%
\begin{align}
&C = - 
\left(\int_0^1 {\rm d}z \int {\rm d}^2 p_{\perp} \, \frac{M^2 z^3}
{\left[{\bf p}_{\perp}^2 + M^2 z^2 + (1-z) m_{\pi}^2\right]^2}\right) \nonumber \\
&\times \left(\int_0^1 {\rm d}z \, z  \int {\rm d}^2 p_{\perp} \, \frac{{\bf p}_{\perp}^2 + M^2 z^2}
{\left[{\bf p}_{\perp}^2 + M^2 z^2 + (1-z) m_{\pi}^2\right]^2}\right)^{-1} \,.
\label{check}
\end{align}
From this relation we see that $-1<C<0$ and cannot be equal to 1, which
verifies the validity of the TM sum rule (\ref{momt}) for the case of ps coupling.

The third point concerns the mixing of operators with opposite chirality 
in the integral equation (\ref{si1}) because of the finite constituent quark mass term in the propagator. 
By noting that the Dirac matrices for massless quark propagators are chiral even, and 
pion-quark couplings
always occur in pairs, we see that for the case of massless quark the chirality of the final product
of Dirac matrices is equal to the chirality of the external quark operators $\gamma^+$, $\gamma^+ \gamma_5$,
$i \sigma^{i+} \gamma_5$. Therefore the term  $\propto \hat{d}_T^{\perp (q \rightarrow Q)} \,
H^{\perp (Q \rightarrow \pi)}$ in the integral equation (\ref{si1}) must arise from
the finite constituent quark mass term in the propagators. The model forms given above actually show that  
$\hat{d}_T^{\perp (q \rightarrow Q)} \propto M^2$, and 
$\hat{h}^{\perp (q \rightarrow Q)} \propto M^2$. Because also
$\hat{h}^{\perp (q \rightarrow \pi)} \propto M$, the integral equation
(\ref{sd3}) gives ${H}^{\perp (q \rightarrow \pi)} \propto M$, and therefore the
second term in the bracket $\left[ \dots \right]$ of (\ref{si1}) is
$\propto M^2$.

In completely the same manner, one can confirm these points also for the case of pv
coupling. First, in order to verify that $-1 < C <0$ from (\ref{y1}), we need the following
3 functions derived from the $q \rightarrow Q$ fragmentation diagram of Fig.2:
\begin{align}
&\tilde{d}^{(q \rightarrow Q)}(z, {\bf p}_{\perp}^2) = \left(\frac{g_A}{2 f_{\pi}} \right)^2 
\frac{1}{4} \frac{1}{(2\pi)^3} \nonumber \\
&\times \left( \frac{1}{1-z} - 
\frac{4 M^2 m_{\pi}^2 \, z (1-z)} 
{\left[{\bf p}_{\perp}^2 + M^2 (1-z)^2 + z m_{\pi}^2 \right]^2} \right) \,,
\label{dqq1}   \\
&\tilde{h}_{T,{\rm neutral}}^{(q \rightarrow Q)}(z, {\bf p}_{\perp}^2) = 
- \tilde{d}_{\rm neutral}^{(q \rightarrow Q)}(z, {\bf p}_{\perp}^2)  \,,
\label{htqq1}  \\
&\tilde{h}_{T,{\rm neutral}}^{\perp (q \rightarrow Q)}(z, {\bf p}_{\perp}^2) = 
\frac{1-z}{2} \left(\frac{g_A}{2 f_{\pi}} \right)^2  \frac{1}{(2\pi)^3}   \nonumber \\
&\times \frac{4 M^4 z^2}
{\left[{\bf p}_{\perp}^2 + M^2 (1-z)^2 + z m_{\pi}^2 \right]^2} \,,
\label{htpqq1}
\end{align}
where the tilde above the functions characterizes the pv coupling, $g_A$ is the
weak axial vector coupling constant on the quark level, and $f_{\pi}$ is the
weak pion decay constant. 
Comparing to the forms (\ref{dqq})-(\ref{htpqq}) for ps coupling, we
see that in pv coupling a kind of contact term appears\cite{ABM05}, and for
a numerical evaluation one needs a scheme which regularizes both the divergencies
of the $z$ integrals and the transverse momentum integrals
\footnote{An example is the invariant mass (or Lepage-Brodsky) regularization scheme
\cite{LB80,BHMY99}.}. Nevertheless, it is straight forward to verify the inequality 
$-1 < C < 0$ on the level of integrands by inserting the above model forms into (\ref{y1}).

Second, the one-pion loop expression for the elementary T-odd function 
$h^{\perp (q \rightarrow \pi)}(z,{\bf p}_{\perp}^2)$ in pv coupling has been given in \cite{ABM05},
and we do not reproduce it here. It has the same prefactor $M \, m_{\pi}$ as in
(\ref{collpi}) of the ps case, 
and from the operator definition (\ref{tplusi}) it follows that 
the function $h^{\perp (q \rightarrow Q)}(z,{\bf p}_{\perp}^2)$ in pv coupling involves 
the same prefactor $M^2$ as in the ps case (\ref{collq}). Together with
the relation (\ref{dq}), which holds also in the pv case, the above discussion
on the mixing of operators with opposite chiralities due to the finite constituent
quark mass term in the propagator holds for pv coupling as well.       

Finally in this Appendix, we list the sum rules for the renormalized functions, including the 
flavor dependence as shown in (\ref{ff1}) and 
(\ref{ff2}):
\begin{align}
&\sum_{\tau_{\pi}} \int_0^1 {\rm d} z \, \int {\rm d}^2 p_{\perp}\,
\hat{f}^{(q \rightarrow \pi)}(z,{\bf p}_{\perp}; {\bf s}) 
= 1\,,  \label{isqp1} \\
&\sum_{\tau_{\pi}} \tau_{\pi} \, \int_0^1 {\rm d} z \, \int {\rm d}^2 p_{\perp}\,
\hat{f}^{(q \rightarrow \pi)}(z,{\bf p}_{\perp}; {\bf s})  
= \frac{2}{3} \tau_q   \,,  \label{ivqp1} \\ 
&\sum_{\tau_Q} \sum_{\pm{\bf S}} \int_0^1 {\rm d} z \, \int {\rm d}^2 p_{\perp}\,
\hat{f}^{(q \rightarrow Q)}(z,{\bf p}_{\perp}; {\bf s}, {\bf S}) 
= 1 \,, \label{isqq1} \\
&\sum_{\tau_Q} \frac{\tau_Q}{2} \, \sum_{\pm {\bf S}}  \int_0^1 {\rm d} z \, \int {\rm d}^2 p_{\perp}\,
\hat{f}^{(q \rightarrow Q)}(z,{\bf p}_{\perp}; {\bf s}, {\bf S})   \nonumber \\
&= - \frac{1}{6} \tau_q \,.    \label{ivqq1} 
\end{align}
Because these sum rules are based only on the normalization condition (\ref{fnorm}) and the flavor
dependence (\ref{ff1}) and (\ref{ff2}), they are model independent. 

\section{Product ansatz and recursion relations}
\setcounter{equation}{0}
We first formulate the product ansatz in terms of the unrenormalized elementary $q \rightarrow Q$
FFs
and the maximum number of pions ($N$) 
which can be produced by the fragmenting quark. Let us denote the first and second terms on the r.h.s. of Eq.(\ref{fg}), 
which correspond
to different hadronic spectator states (namely the vacuum and the one-pion state, respectively)
by $f_v^{(q \rightarrow Q)}$ and $f_p^{(q \rightarrow Q)}$. We use the notations (\ref{multi}) of the main text
to denote multi-dimensional momentum integrations, and also define 
\begin{align}
\left(\sum_{\nu=v,p} \right)^N &\equiv  \sum_{\nu_0=v,p} \, \, \sum_{\nu_1=v,p} \,\, \cdots \,
\sum_{\nu_{N-1}=v,p}    \nonumber   \\
\left(\sum_{\pm {\bf S}_n}\right)^N &\equiv  \sum_{\pm {\bf S}_1} \,  \sum_{\pm {\bf S}_2} \, \cdots \,
\sum_{\pm {\bf S}_N} \,.  \nonumber 
\end{align}
for multiple summations. The basic product ansatz is then as follows:   
\begin{align}
&F^{(q \rightarrow \pi)}(z, {\bf p}_{\perp}; {\bf s})  \nonumber \\
&= \left(\sum_{\nu=v,p}\right)^N  
\sum_{m=1}^N \, \int {\cal D}^N \eta \, \int {\cal D}^{2N} p_{\perp} \, 
\left(\sum_{\pm {\bf S}_n}\right)^N \, \sum_{\tau_{Q_N}}    
\nonumber \\
&\times {f}_{\nu_{0}}^{(q \rightarrow Q_1)}(\eta_1, {\bf p}_{1 \perp}; {\bf S}_1, {\bf s})   \nonumber \\ 
&\times {f}_{\nu_{1}}^{(Q_1 \rightarrow Q_2)}(\eta_2, {\bf p}_{2 \perp} - \eta_2 {\bf p}_{1 \perp}; {\bf S}_2, 
\langle {\bf S}_1 \rangle_{f_{\nu_{1}}})
\times \dots \nonumber \\
&\times \!
{f}_{\nu_{N-1}}^{(Q_{N-1} \rightarrow Q_N)}(\eta_N, {\bf p}_{N \perp} - \eta_N {\bf p}_{N-1 \perp}; {\bf S}_N, 
\langle {\bf S}_{N-1} \rangle_{f_{\nu_{N-1}}}) 
\nonumber \\ 
&\times \delta(z - z_m) \, \delta^{(2)}\left({\bf p}_{\perp} - ({\bf p}_{m-1 \perp} -{\bf p}_{m \perp})\right)
\delta(\nu_{m},1) \nonumber \\
&\times \delta\left(\tau_{\pi}, (\tau_{Q_{m-1}} - \tau_{Q_m}) /2\right) \,.   
\label{ansatz0}
\end{align}
Here the function ${f}_{\nu_i}^{(Q_i\rightarrow Q_j)}(\eta, {\bf p}_{\perp}; {\bf S}_j, {\bf S}_i)$ 
for the $j$th step 
is the unrenormalized elementary FF for the case 
where the incoming quark $(Q_i)$ has zero TM and polarization ${\bf S}_i$ and the
outgoing quark $(Q_j)$ has TM ${\bf p}_{\perp}$ and polarization ${\bf S}_j$. 
The quantities $\langle {\bf S}_i \rangle_{f_{\nu_i}}$ of the $j$th step 
($j=i+1$) denote the average 
polarization of $Q_i$ determined by the functions $f_{\nu_{i-1}}^{(Q_{i-1} \rightarrow Q_i)}$ of 
the $i$th step. 

We now insert the form (\ref{fg}) for each factor $f_{\nu_i}$ of (\ref{ansatz0}) and
sum over the directions of ${\bf S}_j$, where $j=i+1$. As a result, the
factor $\left(1 + {\bf S} \cdot {\bf s}\right)/2$ in $f_v$ of (\ref{fg}) is replaced by unity,
while the spin sum over $\hat{f}$ gives the function (\ref{qqunpol}).   
It is then easy to see that all products with the same number
(call it $k$) of $\hat{f}'s$ and $(N-k)$ number of $Z_Q's$ make the same contribution to $F^{(q\rightarrow \pi)}$.
We therefore can introduce an ordering of the factors in (\ref{ansatz0}),
so that the first $k$ $\eta$'s not equal to one ($\eta_1, \eta_2, \dots \eta_k \neq 1$), and the remaining
$\eta$'s equal to one ($\eta_{k+1}, \eta_{k+2}, \dots \eta_N = 1$), multiply the combinatoric factor
$\left( \begin{array}{cc} 
N \\ k 
\end{array} \right)$ and perform a sum over $k$. For some fixed $k$, only
the terms with $m \leq k$ will contribute to the sum in (\ref{ansatz0}), because $z_m$ in (\ref{zm}) 
must be non-zero. Then Eq.(\ref{ansatz0}) becomes
\begin{align}
&F^{(q \rightarrow \pi)}(z, {\bf p}_{\perp}; {\bf s})   \nonumber \\
&\,\,=  \sum_{m=1}^N \, \sum_{k=m}^N \, P(k) \, 
\int {\cal D}^k \eta \, \int {\cal D}^{2k} p_{\perp} \, \sum_{\tau_{Q_k}}  
\nonumber \\
&\times \hat{f}^{(q \rightarrow Q_1)}(\eta_1, {\bf p}_{1 \perp}; {\bf s})
\nonumber \\ 
&\times \hat{f}^{(Q_1 \rightarrow Q_2)}(\eta_2, {\bf p}_{2 \perp} - \eta_2 {\bf p}_{1 \perp}; \langle {\bf S}_1 \rangle)
\nonumber \\
&\times \cdots \times \hat{f}^{(Q_{k-1} \rightarrow Q_k)}(\eta_k, {\bf p}_{k \perp} - \eta_k {\bf p}_{k-1 \perp};  
\langle {\bf S}_{k-1} \rangle) 
\nonumber \\
&\times \delta(z - z_m) \,\delta^{(2)}\left({\bf p}_{\perp} - ({\bf p}_{m-1 \perp} -{\bf p}_{m \perp})\right)
\nonumber \\
&\times \delta\left(\tau_{\pi}, (\tau_{Q_{m-1}} - \tau_{Q_m}) /2\right) 
\equiv  \sum_{m=1}^N \, F_m^{(q \rightarrow \pi)}(z, {\bf p}_{\perp}; {\bf s}) \,. 
\label{ansatz01}
\end{align}
Here we use the same notation as in the main text for the spin averages, i.e., $\langle {\bf S}_i \rangle$ 
for the $j$th step ($j=i+1$) means the average polarization of $Q_i$ determined by the renormalized function
$\hat{f}^{(Q_{i-1} \rightarrow Q_i)}$ of the $i$th step. The binomial distribution
\begin{align} 
P(k) = \left( \begin{array}{cc} 
N \\ k  \end{array} \right) Z_Q^{N-k} \, \left(1- Z_Q\right)^k  
\label{bino}
\end{align}
is the probability of producing $k$ mesons out of a maximum of $N$ mesons and satisfies the normalization
condition
\begin{align}
\sum_{k=0}^N P(k) = 1 \,.  \label{sump}
\end{align}
For a fixed $m$ in (\ref{ansatz01}), we can integrate over the variables 
$\eta_k$ and ${\bf p}_{k \perp}$ for
$k >m$ by using the normalization (\ref{fnorm}). 
Then, for all $k \geq m$, only the integrations over the same set of variables
$\eta_{\ell}$ and ${\bf p}_{\ell \perp}$ for $\ell = 1,2,\dots m$ remain, and the sum over
$k$ refers only to the probabilities $P(k)$. The integrations over the variables
$\eta_m \,, \, {\bf p}_{m \perp}$ are then performed by using the delta functions. Making a
shift $\eta_m \rightarrow 1 - \eta_m$, and following similar steps as in (\ref{intm}) of the main text, 
the result of these integrations is
\begin{align}
&\sum_{\tau_{Q_m}}\, \int_0^1 {\rm d} \eta_m \, \int {\rm d}^2 p_{m \perp} \,
\,\delta(z - z_m) \nonumber \\ 
&\times \hat{f}^{(Q_{m-1} \rightarrow Q_m)} (\eta_m, {\bf p}_{m \perp} - \eta_m {\bf p}_{m-1 \perp} ; 
 \langle {\bf S}_{m-1} \rangle) \nonumber \\
&\times \delta({\bf p}_{\perp} - ({\bf p}_{m-1 \perp} - {\bf p}_{m \perp}))
\, \delta(\tau_{\pi}, (\tau_{Q_{m-1}} - \tau_{Q_m})/2) 
\nonumber \\
&= \int_0^1 {\rm d} \eta_m \,\delta(z - \eta_1 \eta_2 \dots \eta_m)   \nonumber \\  
&\times \hat{f}^{(Q_{m-1} \rightarrow \pi)} (\eta_m, {\bf p}_{\perp} - \eta_m {\bf p}_{m-1 \perp}; 
\langle {\bf S}_{m-1} \rangle) \,,
\label{intm0}
\end{align}
where the renormalized elementary $q \rightarrow \pi$ splitting function $\hat{f}^{(q \rightarrow \pi)}$ is given by
\begin{align}
\hat{f}^{(q \rightarrow \pi)}(z, {\bf p}_{\perp}; {\bf s}) = \frac{1}{1 - Z_Q} \, 
{f}^{(q \rightarrow \pi)}(z, {\bf p}_{\perp}; {\bf s}) \,.   
\label{fren}
\end{align}
In this way, the function $F_m^{(q \rightarrow \pi)}$ of Eq.(\ref{ansatz01}) becomes
\begin{align}
&F_m^{(q \rightarrow \pi)}(z, {\bf p}_{\perp}; {\bf s}) 
= \left(\sum_{k=m}^N \, P(k)\right) \, 
\int {\cal D}^m \eta \, \int {\cal D}^{2(m-1)} p_{\perp} \nonumber \\
&\! \times \! \hat{f}^{(q \rightarrow Q_1)}(\eta_1, {\bf p}_{1 \perp}; {\bf s}) 
\hat{f}^{(Q_1 \rightarrow Q_2)}(\eta_2, {\bf p}_{2 \perp} - \eta_2 {\bf p}_{1 \perp}; \langle {\bf S}_1 \rangle)
 \cdots  \nonumber \\
&\! \times \! \hat{f}^{(Q_{m-2} \rightarrow Q_{m-1})}(\eta_{m-1}, {\bf p}_{m-1 \perp}\! -\!\eta_{m-1} {\bf p}_{m-2 \perp};  \langle {\bf S}_{m-2} \rangle)  \nonumber \\ 
&\!\times \! \hat{f}^{(Q_{m-1} \rightarrow \pi)} (\eta_m, {\bf p}_{\perp}\! - \! \eta_m {\bf p}_{m-1 \perp}; \langle {\bf S}_{m-1} \rangle)
\delta(z - \eta_1 \eta_2 \cdots \eta_m) \,. 
\label{ansatz02}
\end{align}
In order to obtain a recursion relation for the functions $F_m^{(q\rightarrow \pi)}$, we carry out the following steps
\footnote{The same steps are used in the main text to derive Eq.(\ref{recurs1}), (\ref{neq1}) from (\ref{ansatz2}).}:
First, we make shifts of the
integration variables $\left({\bf p}_{m-1 \perp}, \dots {\bf p}_{1 \perp}\right)  
\rightarrow \left({\bf p}'_{m-1 \perp}, \dots {\bf p}'_{1 \perp}\right)$ according to 
\begin{align}
&{\bf p}_{\ell \perp}' = {\bf p}_{\ell \perp} - \eta_{\ell} \, {\bf p}_{\ell-1 \perp}\,\,\,\,\,\,\,\,\,\,\,\,
(\ell = 1,2,\dots m-1) 
\label{trans}
\end{align}
with ${\bf p}_{0 \perp} \equiv 0$. 
Using these relations recursively, the argument of the function $\hat{f}^{(Q_{m-1} \rightarrow \pi)}$
in (\ref{ansatz02}) becomes
\begin{align}
&{\bf p}_{\perp} - \eta_m \, {\bf p}_{m-1 \perp}   \nonumber \\ 
&= {\bf p}_{\perp} - \eta_m \, {\bf p}'_{m-1 \perp} - \eta_m \eta_{m-1} \, {\bf p}_{m-2}' \dots
\nonumber \\ 
&- \eta_m \eta_{m-1} \cdots \eta_3 \eta_2 \, {\bf p}_{1 \perp}  \,. 
\label{shiftp}
\end{align}
In this way, Eq.(\ref{ansatz02}) can be written as 
\begin{align}
&F_m^{(q \rightarrow \pi)}(z, {\bf p}_{\perp}; {\bf s}) =  \left(\sum_{k=m}^N \, P(k)\right) \, 
\int {\cal D}^m \eta \, \int {\cal D}^{2m} p_{\perp}  \nonumber \\
&\times \hat{f}^{(q \rightarrow Q_1)}(\eta_1, {\bf p}_{1 \perp}; {\bf s})  \nonumber \\
&\times \hat{f}^{(Q_1 \rightarrow Q_2)}(\eta_2, {\bf p}_{2 \perp}; \langle {\bf S}_1 \rangle) \times \cdots  \nonumber \\
&\times  \hat{f}^{(Q_{m-1} \rightarrow \pi)}(\eta_m, {\bf p}_{m \perp}; \langle {\bf S}_{m-1} \rangle) 
\, \delta(z - \eta_1 \eta_2 \cdots \eta_m) \nonumber \\
&\times \delta^{(2)}({\bf p}_{\perp} - {\bf p}_{m \perp} - \eta_m {\bf p}_{m-1 \perp} - 
\eta_m \eta_{m-1} {\bf p}_{m-2 \perp} - \cdots \nonumber \\
&- \eta_m \eta_{m-1} \dots \eta_3 \eta_2 \, {\bf p}_{1 \perp})  \,. 
\label{ansatz03}
\end{align}
Second, we replace $m \rightarrow m-1$ in (\ref{ansatz03}) to obtain an expression for
$F_{m-1}^{(q \rightarrow \pi)}$. In this expression, rename the integration variables
as $\eta_1 \rightarrow \eta_2\,, \eta_2 \rightarrow \eta_3 \,, \dots 
\eta_{m-1} \rightarrow \eta_m$, and similarly for the
TM. Also, rename the quark flavors as $q \rightarrow Q_1,
\, Q_1 \rightarrow Q_2 \,, \dots Q_{m-1} \rightarrow Q_{m}$. 
Third, in the expression (\ref{ansatz03}) for $F_m^{(q \rightarrow \pi)}$, use the following
identities:
\begin{align}
&\delta ( z - \eta_1 \eta_2 \cdots \eta_m) = \int_0^1 {\rm d} \eta \, \delta (z - \eta_1 \eta) \,
\delta (\eta - \eta_2 \eta_3 \cdots \eta_m) \,,  \label{delta1}   \\
&\delta^{(2)} ( {\bf p}_{\perp} - {\bf p}_{m \perp} - \eta_m {\bf p}_{m-1 \perp} - \eta_m \eta_{m-1} {\bf p}_{m-2 \perp} 
\dots \nonumber \\
&- \eta_m \eta_{m-1} \cdots \eta_3 \eta_2 {\bf p}_{1 \perp}) \nonumber \\ 
&= \int {\rm d}^2 {\bf k}_{\perp} \delta^{(2)}( {\bf p}_{\perp} - {\bf k}_{\perp} - \eta {\bf p}_{1 \perp}) 
\nonumber \\
&\times \delta^{(2)} ({\bf k}_{\perp} - {\bf p}_{m \perp} - \eta_m {\bf p}_{m-1 \perp}
- \eta_m \eta_{m-1} {\bf p}_{m-2 \perp} - \dots  \nonumber   \\ 
&- \eta_m \eta_{m-1} \dots - \eta_4 \eta_3 {\bf p}_{2 \perp})   \,. 
\label{delta2}
\end{align}
In (\ref{delta2}) we used $\eta = \eta_2 \eta_3 \cdots \eta_m$ from (\ref{delta1}).

Following the three steps explained above, we obtain the following recursion relation 
for $F_m^{(q \rightarrow \pi)}$:
\begin{align}
&F_m^{(q \rightarrow \pi)} (z, {\bf p}_{\perp}; {\bf s}) \nonumber \\ 
&= R_m \, \int {\cal D}^2 \eta \, \int {\cal D}^{4} p_{\perp} \,  \delta(z - \eta_1 \eta_2)   \nonumber \\
&\times \delta^{(2)}({\bf p}_{\perp} - {\bf p}_{2 \perp} - \eta_2 {\bf p}_{1 \perp})
\hat{f}^{(q \rightarrow Q)}(\eta_1, {\bf p}_{1 \perp} ; {\bf s}) \nonumber \\
&\times F_{m-1}^{(Q \rightarrow \pi)} (\eta_2, {\bf p}_{2 \perp}; \langle {\bf S}_1 \rangle) \,,   
\label{recurs01}
\end{align}
while for $m=1$ we have
\begin{align}
F_1^{(q \rightarrow \pi)} (z, {\bf p}_{\perp}; {\bf s})
= R_1 \, \hat{f}^{(q \rightarrow \pi)} (z, {\bf p}_{\perp}; {\bf s}) \,.
\label{neq01}
\end{align}
The ratios $R_n$ for $n=1,2,\dots N$ are defined as
\begin{align}
R_n = \frac{\sum_{k=n}^N \, P(k)} {\sum_{k=n-1}^N \, P(k)} \,.
\label{rat}
\end{align}
The total FF then becomes
\begin{align}
&F^{(q \rightarrow \pi)} (z, {\bf p}_{\perp}; {\bf s}) = R_1 \, \hat{f}^{(q \rightarrow \pi)} (z, {\bf p}_{\perp}; {\bf s})
\nonumber \\
&+ \sum_{n=2}^N  \, F_n^{(q \rightarrow \pi)} (z, {\bf p}_{\perp}; {\bf s}) \,.
\label{total}
\end{align}
It can be seen from this relation that the sum rules are not satisfied if the maximum number of mesons ($N$)
is finite\cite{IBCTY09}. 
As we explain in the main text, we consider the limit $N \rightarrow \infty$,
where the following relation is satisfied:
\begin{align}
R_n \stackrel{N\rightarrow \infty}{\longrightarrow} 1  \,\,\,\,\,\,\,\,\,\,\,\,(n=1,2, \dots)\,. 
\label{condr}
\end{align}
[We remind that, according to the Moivre-Laplace theorem, in the limit $N\rightarrow \infty$ the 
binomial distribution $P(k)$ of (\ref{bino}) becomes a normal (Gauss)
distribution with the same mean value (equal to $N\left(1-Z_Q\right)$) and variance 
(equal to $N Z_Q \left(1-Z_Q\right)$).]
It then follows from (\ref{total}) and (\ref{recurs01}) that the FF
satisfies the following integral equation in the limit $N\rightarrow \infty$:
\begin{align}
&F^{(q\rightarrow \pi)}(z,{\bf p}_{\perp}; {\bf s}) = \hat{f}^{(q\rightarrow \pi)}(z,{\bf p}_{\perp}; {\bf s})
\nonumber \\
&+ \int {\cal D}^2 \eta \, \int {\cal D}^{4} p_{\perp} \,  \delta(z - \eta_1 \eta_2) \nonumber \\
&\times \delta^{(2)}({\bf p}_{\perp} - {\bf p}_{2 \perp} - \eta_2 {\bf p}_{1 \perp})
\hat{f}^{(q \rightarrow Q)}(\eta_1, {\bf p}_{1 \perp} ; {\bf s})  \nonumber \\
&\times F^{(Q \rightarrow \pi)} (\eta_2, {\bf p}_{2 \perp}; \langle {\bf S}_1 \rangle) \,,  
\label{recurs0}
\end{align}
which is the same as (\ref{recurs}) of the main text.

\section{Mean isospin z-component and transverse momentum of quark remainder}

\setcounter{equation}{0}

In this Appendix we wish to show that, after $N\rightarrow \infty$ fragmentation steps, the mean isospin
$z$-component and the mean TM of the quark remainder are zero. These results
are confirmed in the main part (Sect. III.D.), and for clarity we present alternative
proofs in this Appendix. 

\subsection{Mean isospin z-component of quark remainder}

Denote by $P_N$ the probability that, after $N$ emission of pions, the isospin z-component
of the quark is the same as that of the initial quark. Because in each emission step,
the probability that the quark isospin z-component changes is equal to $2/3$ and that
it does not change is equal to $1/3$, we obtain the recursion relation
\begin{align}
P_{N} = \frac{1}{3} P_{N-1} + \frac{2}{3} \left(1-P_{N-1} \right)
= \frac{2}{3} - \frac{1}{3} P_{N-1} \,. 
\label{recur}
\end{align}
This can be solved with the initial condition $P_0=1$ as
\begin{align}
P_N = \frac{1}{2} \left(1 + \left(- \frac{1}{3} \right)^N \right) \,.
\label{pn}
\end{align}
This shows that in the limit $N\rightarrow \infty$ $P_N$ becomes $1/2$, i.e., that quark
remainder has equal probabilities for isospin z-component $\pm 1/2$, and therefore its
mean isospin z-component must be zero. More explicitly, if $\tau_q/2$ is the isospin z-component
of the initial quark, then after $N$ emission steps the quark has average isospin z-component
\begin{align}
\frac{\tau_q}{2} P_N - \frac{\tau_q}{2} \left(1-P_N\right) = \frac{\tau_q}{2} \left(2 P_N -1\right)
= \frac{\tau_q}{2} \left(- \frac{1}{3} \right)^N \,, 
\label{in}
\end{align}
which vanishes in the limit $N\rightarrow \infty$.

\subsection{Mean TM of quark remainder}

\setcounter{equation}{0}

According to our product ansatz (III.6), the probability for a fragmentation chain is given by
the products of elementary $q\rightarrow Q$ splitting functions. The delta-functions in
(III.6) select a meson which is produced in the $m$-th step, and the summation over $m$ gives the
probability for semi-inclusive pion production. Instead of selecting the pions, we now select
the final quark by the delta functions. Because we are interested in the isoscalar case, we sum over the flavors of the
final quark. This gives for the probability density of $q \rightarrow Q_N$
\begin{align}
&P(z, {\bf p}_{\perp}; {\bf s}) = {\rm lim}_{N\rightarrow \infty} \int {\cal D}^N \eta
\int {\cal D}^{2N} p_{\perp} \sum_{\tau_{Q_N}}  \nonumber \\
& \times \hat{f}^{(q \rightarrow Q_1)}(\eta_1, {\bf p}_{1 \perp}; {\bf s})  
\hat{f}^{(Q_1 \rightarrow Q_2)} (\eta_2, {\bf p}_{2 \perp} - \eta_2 {\bf p}_{1 \perp}; \langle {\bf S}_1\rangle)      
\nonumber \\
&\times \dots \times \hat{f}^{(Q_{N-1} \rightarrow Q_{N})} (\eta_N, {\bf p}_{N \perp} - \eta_N {\bf p}_{N-1 \perp}; 
\langle {\bf S}_{N-1} \rangle)   \nonumber  \\
&\times \delta(z-\eta_N) \delta^{(2)}({\bf p}_{\perp} - {\bf p}_{N \perp}) \,. 
\label{prodf}
\end{align}
Because each factor has the flavor dependence (\ref{isivqq}), it is easy to see that,
after the flavor summations, all elementary functions should be replaced by the isoscalar functions
$\hat{f}_{(0)}^{(q \rightarrow Q)}$.
The mean TM of the quark remainder is obtained by multiplying (\ref{prodf}) by
${\bf p}_{\perp}$ and integrating over $z$ and ${\bf p}_{\perp}$. This gives  
\begin{align}
&\langle {\bf p}_{\perp} \rangle_{\rm rem} \equiv 
\int {\rm d}^2 p_{\perp} \, {\bf p}_{\perp}  \int_0^1 {\rm d} z \, P(z, {\bf p}_{\perp}; {\bf s})
\nonumber \\
&= {\rm lim}_{N\rightarrow \infty} \int {\cal D}^N \eta  
\int {\cal D}^{2N} p_{\perp} \nonumber \\
& \times \hat{f} (\eta_1, {\bf p}_{1 \perp}; {\bf s}) \,\,  
\hat{f} (\eta_2, {\bf p}_{2 \perp} - \eta_2 {\bf p}_{1 \perp}; \langle {\bf S}_1\rangle)      
\nonumber \\
&\times \dots \times \hat{f} (\eta_N, {\bf p}_{N \perp} - \eta_N {\bf p}_{N-1 \perp}; 
\langle {\bf S}_{N-1} \rangle) \, {\bf p}_{N \perp}  \,,  
\label{prod1}
\end{align}
where now all functions in the product refer to the isoscalar part of the elementary $q \rightarrow Q$ splitting function.
Next we use the shifts of integration variables (\ref{trans}) for all $\ell = 1,2,\dots N$. 
Using these relations recursively, as explained in (\ref{shiftp}), to express ${\bf p}_{N \perp}$ by the new variables, 
we obtain 
\begin{align}
&\langle {\bf p}_{\perp} \rangle_{\rm rem}
= {\rm lim}_{N\rightarrow \infty} \int {\cal D}^N \eta
\int {\cal D}^{2N} p_{\perp} \, \hat{f} (\eta_1, {\bf p}_{1 \perp}; {\bf s}) \nonumber \\   
&\times \hat{f} (\eta_2, {\bf p}_{2 \perp}; \langle {\bf S}_1\rangle) \times \dots \times 
\hat{f} (\eta_N, {\bf p}_{N \perp}; \langle {\bf S}_{N-1} \rangle) \nonumber \\ 
&\times \left({\bf p}_{N \perp} + \eta_N {\bf p}_{N-1 \perp} + \eta_N \eta_{N-1} {\bf p}_{N-2 \perp}
+ \dots \right. \nonumber  \\
&\left. + \eta_N \eta_{N-1} \dots \eta_2 {\bf p}_{1 \perp} \right) \,. 
\label{prod2}
\end{align}
Remember that the function for the $n$th step in this product has the form (\ref{qqunpol}) 
\begin{align}
&\hat{f} (\eta_n, {\bf p}_{n \perp}; \langle {\bf S}_{n-1} \rangle) \nonumber \\
&= 2 \left[ \hat{d} (\eta_n, {\bf p}_{n \perp}^2) + \frac{1}{M \eta_n} 
\left({\bf p}_{n \perp} \times \langle {\bf S}_{ n-1} \rangle \right)^3 \, 
\hat{h}^{\perp} (\eta_n, {\bf p}_{n \perp}^2) \right]   
\label{drive}
\end{align}
with $\langle {\bf S}_0\rangle \equiv {\bf s}$. Also, remember that for $\langle {\bf S}_{ n} \rangle$ in the 
function for the $(n+1)$ step we have the recursion relation
(see Eq. (\ref{sav}))

\begin{align}
&\langle {\bf S}_{ n} \rangle \cdot \hat{f}(\eta_{n}, {\bf p}_{n \perp}; \langle {\bf S}_{ n-1} \rangle)
\nonumber \\
& =   2 \left[ \frac{1}{M \eta_n} {\bf p}'_{n \perp} \, \, \hat{d}_T^{\perp} (\eta_n, {\bf p}_{n \perp}^2) 
+ \langle {\bf S}_{ n-1} \rangle \hat{h}_T (\eta_n, {\bf p}_{n \perp}^2) \right.  \nonumber \\
&+ \left. \frac{1}{M^2 \eta_n^2} 
{\bf p}_{n \perp} \left(\langle {\bf S}_{ n-1} \rangle \cdot {\bf p}_{n \perp} \right) 
\hat{h}_T^{\perp} (\eta_n, {\bf p}_{n \perp}^2) \right] \,,   
\label{savf}
\end{align}
where the vector ${\bf p}_{\perp}'$ is defined by ${\bf p}_{\perp}' = 
\left(-p_{\perp}^2, p_{\perp}^1 \right)$ if ${\bf p}_{\perp} = (p_{\perp}^1, p_{\perp}^2)$. 
(We also remind that longitudinal quark polarizations do not contribute
to inclusive pion production, and therefore all spin vectors of this Appendix can be replaced
by their transverse parts.)  

Consider now the integral over $\left( \eta_N, {\bf p}_{N \perp}\right)$ in the second term of 
$(\dots)$ in (\ref{prod2}). Here
only the spin independent term $\propto \hat{d}$ of the $N$th factor in the product (\ref{prod2})
contributes, which gives the 
longitudinal momentum fraction left to the quark in one step. We denote this by $K$, where
clearly $K<1$. For example, using the model forms of Appendix C for the case of ps coupling,
we have
\begin{align}
&K \equiv 2 \int_0^1 {\rm d} \eta \, \eta \int {\rm d}^2 p_{\perp} \, \hat{d}(\eta, {\bf p}_{\perp}^2)
\nonumber \\ 
&= \left(\int_0^1 {\rm d} \eta \, \eta (1-\eta)
\int{\rm d}^2 p_{\perp} \frac{{\bf p}_{\perp}^2 + M^2 \eta^2}
{\left[{\bf p}_{\perp}^2 + M^2 \eta^2 + (1-\eta) m_{\pi}^2 \right]^2} \right) \nonumber \\
&\times \left(\int_0^1 {\rm d}\eta \, \eta  \int {\rm d}^2 p_{\perp} \, \frac{{\bf p}_{\perp}^2 + M^2 \eta^2}
{\left[{\bf p}_{\perp}^2 + M^2 \eta^2 + (1-\eta) m_{\pi}^2\right]^2} \right)^{-1} \,.
\label{k}
\end{align}
For the third term in $(\dots)$ of (\ref{prod2}) we can
carry out the integrations over $\left(\eta_N, {\bf p}_{N \perp}\right)$ and 
$\left(\eta_{N-1}, {\bf p}_{N-1 \perp}\right)$ to get a factor $K^2$, and
so on. Therefore Eq.(\ref{prod2}) can be written as
\begin{align}
&\langle {\bf p}_{\perp} \rangle_{\rm rem} 
= {\rm lim}_{N\rightarrow \infty} \, \sum_{n=1}^N \, {\bf I}_n \, K^{N-n}  \,, 
\label{last1}
\end{align}
where we defined the integrals ${\bf I}_n$ by
\begin{align}
{\bf I}_n &= \int {\cal D}^n \eta  \int {\cal D}^{2n} p_{\perp}  \,\, 
\hat{f} (\eta_1, {\bf p}_{1 \perp}; {\bf s}) \,\,   
\hat{f} (\eta_2, {\bf p}_{2 \perp}; \langle {\bf S}_1\rangle)  \nonumber \\
&\times \dots \times 
\hat{f} (\eta_n, {\bf p}_{n \perp}; \langle {\bf S}_{n-1} \rangle) \, {\bf p}_{n \perp} \,.
\end{align}  
These integrals can be evaluated in closed form by using (\ref{drive}) and (\ref{savf}).
The result is
\begin{align}
I_n^i = - \left( \epsilon^{ij}  \, s_j \right) A \cdot C^{n-1}  \,, 
\label{ini}
\end{align}
where we defined the constant $A$ by
\begin{align}
A = \int_0^1 {\rm d}\eta \int {\rm d}^2 p_{\perp} \, \frac{{\bf p}_{\perp}^2}{M \eta} \, 
h^{\perp}(\eta, {\bf p}_{\perp}^2) \,.
\label{adef}
\end{align} 
The constant $C$ was defined already in (\ref{y}), where it was shown that $|C|<1$, and
that $C$ has the physical meaning of the quark depolarization factor for one fragmentation step.  
The TM of the quark remainder is then
finally obtained from (\ref{last1}) as
\begin{align}
&\langle {p}^i_{\perp} \rangle_{\rm rem} = - \left( \epsilon^{ij}  \, s_j \right) \,A \,
\lim_{N \rightarrow \infty} \frac{K^N - C^N}{K-C} = 0 \,,  
\label{last}
\end{align}
where we used $|K|<1$ and $|C|<1$. 
We finally note that for the elementary process the average TM of the final quark is
given by ${\bf I}_1 \propto A$, which is nonzero. It is only after an infinite
chain of fragmentation processes that the average TM of the final quark becomes zero.
As we noted already in the main text, the magnitude of the fluctuation
$\sqrt{\langle {\bf p}^2_{\perp} \rangle_{\rm rem}}$ is nonzero.

\end{document}